\documentclass{emulateapj}
\pdfoutput=0
\shorttitle{Kepler-419 in 3D}
\shortauthors{Dawson et al.}
\usepackage{natbib}
\usepackage{amsmath}
\usepackage{longtable}
\usepackage{graphicx}
\def\rhocirc{\rho_{\rm circ}}
\def\kep{\emph{Kepler\ }}
\def\afinal{a_{\rm final}}
\def\teff{T_{\rm eff}}
\def\feh{\rm[Fe/H]\ }

\begin{document}
\slugcomment{Submitted to ApJ on March 12, 2014. Accepted on June 12, 2014.}
\title{Large eccentricity, low mutual inclination: the three-dimensional architecture of a hierarchical system of giant planets}
\author{Rebekah I. Dawson\altaffilmark{1,2}}
\author{John Asher Johnson\altaffilmark{3}}
\author{Daniel C. Fabrycky\altaffilmark{4}}
\author{Daniel Foreman-Mackey\altaffilmark{5}}
\author{Ruth A. Murray-Clay\altaffilmark{3}}
\author{Lars A. Buchhave\altaffilmark{6,7}}
\author{Phillip A. Cargile \altaffilmark{8}}
\author{Kelsey I. Clubb\altaffilmark{1} }
\author{Benjamin J. Fulton\altaffilmark{9} }
\author{Leslie Hebb\altaffilmark{10}}
\author{Andrew W. Howard\altaffilmark{9}}
\author{Daniel Huber\altaffilmark{11,12}}
\author{Avi Shporer\altaffilmark{13,14,16}  }
\author{Jeff A. Valenti \altaffilmark{15}}
\altaffiltext{1}{{Department of Astronomy, University of California, Berkeley, Hearst Field Annex B-20, Berkeley CA 94720-3411}}
\altaffiltext{2}{{{\tt  rdawson@berkeley.edu}; Miller Fellow}}
\altaffiltext{3}{Harvard-Smithsonian Center for Astrophysics; Institute for Theory and Computation, 60 Garden St, MS-51, Cambridge, MA 02138}
\altaffiltext{4}{Department of Astronomy and Astrophysics, University of Chicago, 5640 S. Ellis Ave, Chicago, IL 95064}
\altaffiltext{5}{Center for Cosmology and Particle Physics, Department of Physics, New York University, Washington Place, New York, NY, 10003, USA}
\altaffiltext{6}{Niels Bohr Institute, University of Copenhagen, DK-2100, 
		Copenhagen, Denmark}
\altaffiltext{7}{Centre for Star and Planet Formation, Natural History Museum of 
		Denmark, University of Copenhagen, DK-1350, Copenhagen, Denmark}
\altaffiltext{8}{Department of Physics and Astronomy, Vanderbilt University, Nashville, TN 37235, USA}
\altaffiltext{9}{Institute for Astronomy, University of Hawaii, 2680 Woodlawn Drive, Honolulu, HI 96822-1839 }
\altaffiltext{10}{Department of Physics, Hobart and William Smith Colleges, Geneva, NY 14456, USA}
\altaffiltext{11}{NASA Ames Research Center, Moffett Field, CA 94035, USA}
\altaffiltext{12}{SETI Institute, 189 Bernardo Avenue, Mountain View, CA 94043, USA}
\altaffiltext{13}{Division of Geological and Planetary Sciences, California Institute of Technology, Pasadena, CA 91125}
\altaffiltext{14}{Jet Propulsion Laboratory, California Institute of Technology, 4800 Oak Grove Drive, Pasadena, CA 91109, USA}
\altaffiltext{15}{Space Telescope Science Institute, 3700 San Martin Drive, Baltimore, MD 21218, USA}
\altaffiltext{16}{Sagan Fellow}

\begin{abstract}
We establish the three-dimensional architecture of the Kepler-419 (previously KOI-1474) system to be eccentric { yet with a low mutual inclination}. Kepler-419b is a warm Jupiter at semi-major axis $a = 0.370^{+0.007}_{-0.006}$ AU with a large eccentricity (e=0.85$^{+0.08}_{-0.07}$) measured via the ``photoeccentric effect.'' It exhibits transit timing variations induced by the non-transiting Kepler-419c, which we uniquely constrain to be a moderately eccentric ($e=0.184\pm0.002$), hierarchically-separated ($a=1.68\pm0.03$ AU) giant planet ($7.3\pm0.4M_{\rm~Jup}$). We combine sixteen quarters of \kep photometry, radial-velocity (RV) measurements from the HIgh Resolution Echelle Spectrometer (HIRES) on Keck, and improved stellar parameters that we derive from spectroscopy and asteroseismology. From the RVs, we measure the mass of inner planet to be $2.5\pm0.3M_{\rm~Jup}$ and confirm its photometrically-measured eccentricity, refining the value to e=0.83$\pm$0.01. The RV acceleration is consistent with the properties of the outer planet derived from TTVs. We find that, despite their sizable eccentricities, the planets are coplanar to within $9^{+8}_{-6}$ degrees, and therefore the inner planet's large eccentricity and close-in orbit are unlikely to be the result of Kozai migration. Moreover, even over many secular cycles, the inner planet's periapse is most likely never small enough for tidal circularization. Finally, we present and measure a transit time and impact parameter from four simultaneous ground-based light curves from 1m-class telescopes, demonstrating the feasibility of ground-based follow-up of \kep giant planets exhibiting large TTVs. 
\end{abstract}

\section{Introduction}
In systems of giant planets, we might expect highly eccentric orbits to go hand-in-hand with large mutual inclinations. The large obliquities between many hot Jupiters and their host stars are interpreted as signatures of the multi-body gravitational processes that led to the hot Jupiter achieving its close-in orbit through high eccentricity migration \citep{2010W,2011M,2012AW,2012N}. Conversely, many of the multi-body interactions that trigger high eccentricity migration, require large mutual inclinations (the Kozai mechanism; e.g. \citealt{2003W,2011NF}), or are likely to produce them (planet-planet scattering, e.g. \citealt{1996R}, with mutual inclinations explored by \citealt{2008C}; secular chaos, \citealt{2011WL}). Besides being a channel for hot Jupiters, including the subset with { orbits misaligned with the host stars' spin axis}, the dynamical interactions that lead to high eccentric migration are also successful in producing the wide distribution of giant planet eccentricities from multi-planet systems with initially circular orbits (e.g. \citealt{2008J,2013KR}). Moreover, even systems of eccentric bodies that are initially flat can exchange angular momentum to achieve large mutual inclinations (e.g. \citealt{2014L}). However, in contrast to the large number of hot Jupiter obliquity measurements (see \citealt{2012AW} and references therein), only a small collection of systems of giant planets have measured mutual inclinations. Most are composed of planets on co-planar, low eccentricity, resonant orbits (i.e. GJ 876, \citealt{2010RL}; Kepler-30, \citealt{2012SFW}; KOI-872, \citealt{2012NK}; Kepler-56, \citealt{2013HC}) that are likely products of disk migration (e.g. \citealt{1980G}). The exception published to date is upsilon Andromedae \citep{2010M}, a hierarchical system with a mutual inclination of 30$^\circ$ measured from astrometry of the host star. Here we investigate the three-dimensional architecture of the hierarchical Kepler-419 (previously KOI-1474) system, which hosts a highly eccentric warm Jupiter and widely-separated perturbing body (\citealt{2012DJM}, D12 hereafter), exactly the sort of system for which violent dynamical histories are posited and a large mutual inclination is, depending on the mechanism, either common or necessary (see above). 

The  \kep candidate KOI-1474.01 (Kepler-419b hereafter, as we will confirm its planetary nature here) came to our attention during our search for hot Jupiters' posited progenitors: Jupiters on highly-eccentric orbits that are migrating via tidal friction. \citet{2012SKDT} argued that if hot Jupiters are produced via multi-body interactions, not disk migration, the \kep sample should contain half a dozen super-eccentric planets with periapses within $\sim$0.05 AU. A super-eccentric proto-hot Jupiter is a planet caught in the act of migrating from a wide, eccentric orbit to a close-in, circular orbit. Among the planet candidates whose eccentricities we measured via the ``photoeccentric effect" \citep{2012DJ}, we found an overall lack of proto-hot Jupiters compared to the theoretical expectation \citep{2012DMJ}. However, the most likely to be a proto-hot Jupiter was Kepler-419b, a warm Jupiter (D12). { We found Kepler-419b to be highly eccentric (e=0.85$^{+0.08}_{-0.07}$), with a} final orbital period, if it were to undergo complete tidal circularization with no change in angular momentum, { of} $ P_{\rm final} = P (1-e^2)^{3/2} = 14^{+9}_{-10}$ days. Furthermore, it exhibits large { transit} time variations (on the order of an hour) caused by a non-transiting companion, possibly the ``smoking gun'' that caused the inner planet's eccentric, close-in orbit. Our original analysis -- based on eight quarters of \kep data -- left two open questions. First, is the inner planet's periapse actually close enough to the star for the planet to undergo significant tidal circularization over the star's lifetime, or is the planet a failed-hot Jupiter? Second, what is the mass and mutual inclination of the outer companion and what do its properties imply about the system's dynamical history? With sixteen quarters of \kep data and high precision RV measurements from Keck HIRES spanning a year, we now seek to address these questions.

In addition to hosting a highly-eccentric warm Jupiter, the Kepler-419 system is special because of the possibility of extracting the properties of the non-transiting planet from the TTVs without the degeneracies that often arise. When TTVs are caused by a perturber near orbital resonance, the period of the TTVs depends on how close the planets are to perfect commensurability and therefore the TTV period itself does not uniquely constrain which orbital resonance the planets are near. Consequently if the perturber is non-transiting, it is often not possible to uniquely determine its mass and orbital period (e.g. as was the case for the first non-transiting planet discovered through TTVs by \citealt{2011B}), though sometimes {
 transit duration variations (TDVs)} can allow one to distinguish (e.g. \citealt{2012NK}). TTVs caused by proximity to orbital resonance are also plagued by a degeneracy between planetary mass and eccentricity, which can only be broken if the planets have zero free eccentricity \citep{2012LX,2013WL}.  In contrast, the TTVs of Kepler-419b are not caused by orbital resonance but by the change in the gravitational potential over the { orbital timescale of the outer, perturbing planet} (see \citealt{2003BE}; Borkovits et al. 2003; \citealt{2005A}, Section 4; D12, Section 5), so there is no degeneracy in the non-transiting planet's orbital period. Moreover, the shape of the TTV signal --- well constrained by Kepler-419b's large (hour) amplitude TTVs (signal-to-noise ratio of order 100) --- allows us to uniquely determine the perturbing planet's mass, eccentricity, and mutual inclination, yielding a more complete set of dynamical information than available for most observed planetary systems.

{ To characterize the Kepler-419 system in detail, we combine information derived from the transit light curves; TTV and RV measurements; and improved host star characterization. In Section \ref{sec:star}, we present updated parameters for the host star based on spectroscopy and asteroseismology. In Section \ref{sec:lc}, we measure the eccentricity of the inner planet and its TTVs from the transit light curves. In Section \ref{sec:ground}, we measure transit times from four ground-based light curves, demonstrating the feasibility of ground-based follow-up of \kep giant planets exhibiting TTVs.  In Section \ref{sec:RVs}, we present high-precision HIRES RV measurements that confirm that the transiting object is planetary mass and has an eccentricity in agreement with the value we measured using the photoeccentric effect. { The measurements are also} consistent with the acceleration expected from the outer planet. In Section \ref{sec:threed}, we derive constraints on the system's three-dimensional architecture from the TTVs, revealing the perturber to be a planet-mass, nearly co-planar object located at $1.68\pm0.03$ AU. In Section \ref{sec:mig}, we discuss which dynamical histories and migration scenarios are consistent with the system's current configuration. We { summarize our conclusions} in Section \ref{sec:summary}.}

\section{Improved stellar characterization}
\label{sec:star}

{ In D12, we characterized host star Kepler-419 as a rapidly-rotating, main-sequence F star with a temperature near that at which stars transition from having outer convective envelopes to having fully radiative outer layers.} Here we present improved host star properties that we will use to better characterize the planets in the system. In Section \ref{sec:sme}, we derive properties from high-resolution spectroscopy with Keck HIRES. In Section \ref{sec:astero}, we show that these properties are consistent with upper-limits based on the asteroseismological non-detection. 

\subsection{Host star properties derived from spectroscopy}
\label{sec:sme}

{ Previously (D12), we presented stellar properties (effective temperature, surface gravity, metallicity, and projected rotational velocity) derived from two high-resolution HIRES spectra. Here we conduct identical observations with a longer exposure time of 2400 seconds to obtain  S/N $\approx 120$ at 550 nm}. Here we analyze { the new spectrum} using three approaches, { the second two of which are detailed in Appendix \ref{app:spec}}. The first is a pipeline (Phillip Cargile,~Leslie Hebb,~et~al.~2014,~in~preparation) that calls Spectroscopy Made Easy (SME; \citealt{1996V,2005V}) hundreds of times to assess parameter covariances and sensitivity to initial conditions. The pipeline uses an expanded line list relative to the \citet{1996V} version, who analyzed higher { S/N} spectra of stars cooler than Kepler-419.  In column 1, section 1 of Table \ref{tab:star}, we list the stellar properties and their formal uncertainties. Based on comparison with stars from \citet{2005V}, \citet{2012T}, and \citet{2013H}, Cargile,~Hebb,~et~al. derived systematic uncertainties of 69 K, 0.10 dex, 0.07 dex, and 1.3 km/s in the stellar effective temperature, surface gravity, metallicity, and projected rotation speed; these systematic uncertainties are added in quadrature to the formal uncertainties. 
 
\begin{deluxetable*}{llll}
\tabletypesize{\footnotesize}%
\tablecaption{Stellar Parameters for KOI 1474 \label{tab:star}}
\tablewidth{0pt}
\tablehead{
\colhead{Parameter}    & \colhead{From spectrum\tablenotemark{ a} }& \colhead{From model\tablenotemark{ a} }& \colhead{With asteroseismology limits\tablenotemark{ a} }}
\startdata
Cargile,~Hebb,~et~al. pipeline (default throughout paper)\\
Projected rotation speed, $v_{\rm rot} \sin i_s$ [km s$^{-1}$] 			& 14.41$\pm $1.3							\\
Stellar effective temperature, $\teff$ [K] 							& 6430$\pm$79 		&6422$^{+75}_{-79}$ 		&6421$^{+76}_{-80}$ \\
Iron abundance, \feh											&0.176 $\pm$0.07		&0.16$^{+0.08}_{-0.04}$ 		&0.16$^{+0.08}_{-0.04}$ 	\\
Surface gravity, $\log g [$cms$^{-2}$] 							&4.10$\pm$0.12		&4.16$^{+0.11}_{-0.14}$ 		&4.19$^{+0.09}_{-0.09}$ \\
Stellar mass, $M_{\star}$ [$M_{\odot}$]							&					&1.42$^{+0.12}_{-0.08}$ 		&1.40$^{+0.06}_{-0.08}$\\
Stellar radius, $R_{\star}$ [$R_{\odot}$] 							&					&1.64$^{+0.35}_{-0.24}$ 		&1.57$^{+0.20}_{-0.18}$ \\
Stellar density, $\rho_{\star}$ [$\rho_{\odot}$]						&					& 0.32$^{+0.16}_{-0.13}$		& 0.36$^{+0.14}_{-0.10}$\\
\hline	
Valenti et al. SME, Version 288\\
Projected rotation speed, $v_{\rm rot} \sin i_s$ [km s$^{-1}$] 			& 14$\pm $0.44										\\
Stellar effective temperature, $\teff$ [K] 							& 6463$\pm235$ 		&6357$\pm$230			&6341$\pm$230 \\
Iron abundance, \feh											&0.14$\pm0.09$ 		&0.12$^{+0.12}_{-0.08}$ 		&0.12$^{+0.12}_{-0.08}$ 	\\
Surface gravity, $\log g [$cms$^{-2}$] 							&4.3$\pm0.28$ 		&4.25$^{+0.08}_{-0.16}$ 		&4.27$^{+0.08}_{-0.11}$ \\
Stellar mass, $M_{\star}$ [$M_{\odot}$]							&					&1.32$^{+0.16}_{-0.12}$ 		&1.32$^{+0.12}_{-0.10}$\\
Stellar radius, $R_{\star}$ [$R_{\odot}$] 							&					&1.42$^{+0.36}_{-0.19}$ 		&1.39$^{+0.25}_{-0.17}$ \\
Stellar density, $\rho_{\star}$ [$\rho_{\odot}$]						&					& 0.46$^{+0.20}_{-0.20}$		& 0.49$^{+0.18}_{-0.17}$\\		
\hline
SPC	\\
Projected rotation speed, $v_{\rm rot} \sin i_s$ [km s$^{-1}$] 			& 14.9$\pm $0.5							\\
Stellar effective temperature, $\teff$ [K] 							& 6376$\pm$77	 	&6369$\pm$79 		&6362$^{+82}_{-80}$\\
Metal abundance, [m/H]											&0.11$\pm$0.10 		&$0.12^{+0.08}_{-0.12}$ 	&0.12$^{+0.08}_{-0.12}$ 	\\
Surface gravity, $\log g [$cms$^{-2}$] 							&4.06$\pm$0.10		&4.08$\pm0.11$ 		&4.14$^{+0.09}_{-0.07}$  \\
Stellar mass, $M_{\star}$ [$M_{\odot}$]							&					&1.42$^{+0.12}_{-0.09}$	&1.38$\pm$ 0.08\\
Stellar radius, $R_{\star}$ [$R_{\odot}$] 							&					&1.78$^{+0.29}_{-0.26}$ 	&1.65$^{+0.16}_{-0.19}$ \\
Stellar density, $\rho_{\star}$ [$\rho_{\odot}$]						&					& 0.25$^{+0.13}_{-0.08}$	& 0.30$^{+0.12}_{-0.06}$	\\			

\enddata
\tablenotetext{a}{The uncertainties represent the 68.3\% confidence interval of the posterior distribution.}
\end{deluxetable*}

We use the approach described by D12 to fit the observed stellar properties using the \citet{2007T} stellar evolution models, except that we do not impose priors from TRILEGAL (TRIdimensional modeL of thE GALaxy; \citealt{2005G}) on the stellar parameters (note that the priors in D12 had no detectable effect on the posteriors). We list the derived mass, radius, and density in column 2 of Table \ref{tab:star}. The stellar parameters obtained using the three spectroscopic analyses are consistent within their uncertainties.

\subsection{Upper-limits from asteroseismology}
\label{sec:astero}
In the six quarters of short cadence data (Q9-Q14), we do not detect p-mode oscillations at the expected frequencies. Because the amplitude of these oscillations increases with stellar radius, we can place an upper limit on the radius of host star Kepler-419. We calculate the detection probability using the method by \citet{2011C}, assuming 550d of short-cadence data and the shot noise expected for a star of Kepler-419's magnitude. At the 99\% confidence level, we find the non-detection sets limits of $\log g >$ 4 and $R_\star < 1.9 R_\odot$ (i.e. both criteria must be met), consistent with the spectroscopic solution. This strict lower limit on $\log g$ is a consequence of the oscillation amplitude changing rapidly as a function of $\log g$ at the location of Kepler-419 on the Hertzsprung-Russell diagram. A star with a lower surface gravity or larger radius than the thresholds derived here would have a clear signal, as we tested by injecting artificial solar-like oscillations into the data for a several representative model stars. For example, a star with $\teff=6200$ K and $\log g=3.95$ has an easily detectable signal. 

Our lower limit on $\log g$ compares well with the results by \citet{2014CC}, who find $\log g > 3.92 \pm 0.04$. The slightly more conservative limit by \citet{2014CC} is likely the result of using a higher effective temperature derived from broadband photometry \citep[6743K,][]{2012P}, which is known to decrease the expected oscillation amplitude \citep{2011C} and hence result in a less stringent upper limit on $\log g$. Since the temperature in our study is based on a high-resolution spectrum rather than a broadband color (which is susceptible to reddening), we adopt the more optimistic lower limit as our final estimate.

We repeat our derivation of the stellar properties but impose this upper limit on stellar radius and lower limit on surface gravity. The resulting parameters are listed in Column 4 of Table \ref{tab:star}. The effect on the stellar parameters is weak, for example changing the stellar density from 0.32$^{+0.16}_{-0.13}$ to 0.36$^{+0.14}_{-0.10}$ solar. We will use the stellar density posterior in our photometric measurement of the transiting planet's eccentricity in Section \ref{sec:lc}, and we will use the stellar mass and radius posteriors as priors in our fits throughout this paper.

\section{Orbital properties of the transiting planet from transit light curves}
\label{sec:lc}

Here we present updated properties of the transiting planet Kepler-419b based on sixteen quarters of \kep data. We measure its eccentricity, TTVs, and an impact parameter for each light curve. In this section, we fit and detrend the transit light curves using, for comparison, two different approaches.

In the first approach, we perform initial detrending before fitting the light curves, described in Appendix \ref{app:fits}.  Next we follow D12 (Section 5.1) to fit the transit light curves using the Markov Chain Monte Carlo (MCMC) fitting procedure, with the CW09 wavelet likelihood and \citet{2002M} light curve model, in the Transit Analysis Package ({\tt TAP}; \citealt{2011G}). The CW09 wavelet likelihood includes two noise parameters: a normalized white noise parameter, $\sigma_w$ and a normalized red noise parameter, $\sigma_r$. The red noise parameter accounts for correlated noise caused by stellar or instrumental variations. { Updated from D12,} we have modified the {\tt TAP} package to fit for impact parameter $b$ instead of inclination, $\rhocirc$ instead of $a/R_\star$, and the limb darkening coefficients $q_1$ and $q_2$ recommended by \citet{2013K}, which are related to the traditional quadratic limb darkening coefficients $\mu_1$ and $\mu_2$ used in the \citet{2002M} light curve model  by $q_1=(\mu_1+\mu_2)^2$ and $q_2 = 0.5 \mu_1 (\mu_1+\mu_2)^{-1}$ . We make a correction\footnote{In the definition of the jump probability in {\tt TAP}, there is an extraneous factor of 2. The necessary factor of 2 is already present in the CW09 likelihood function. Note that this correction was implemented, although not mentioned, in \citet{2012DJ}, D12, and \citet{2012DMJ}} to the likelihood function. Whereas only long cadence data was available for the fits in D12, \kep observed the star in short cadence in Q9-Q16, yielding ten short cadence transits. Therefore we add a separate pair of noise parameters that characterize the short-cadence light curves. (We also perform fits, not tabulated here, in which we allow the noise parameters to be different for each light curve, but we find that the parameters and their uncertainties are essentially identical.) To better account for uncertainties in the detrending, we allow each light curve to have two additional free parameters: a slope and intercept for a linear trend. Finally, we allow $b$ to be a free parameter for each light curve, allowing for the transit duration variations that could occur if the perturber is mutually inclined. We report the planetary parameters derived from the transit light curves in Tables \ref{tab:lc}, \ref{tab:lcttv}, and \ref{tab:lcb}. 

In our second approach, we use a more flexible noise model---designed to capture both the standard measurement uncertainties and longer term systematics or trends---and directly model the standard aperture photometry (SAP) flux without detrending. The noise is modeled as a Gaussian process \citep[see, for example,][]{rasmussen} for which the elements of the covariance matrix $K$ are given by the function
\begin{eqnarray}
K_{ij} &=& [\sigma_i^2 + s^2]\,\delta_{ij} +
            \alpha^2\,\exp \left ( -\frac{{[t_i - t_j]}^2}{2\,\tau^2} \right )
\end{eqnarray}
\noindent for which $\sigma_i$ is the observational uncertainty on data point $i$ (we fix this value to that reported by the \kep pipeline), $\delta_{ij}$ is the Kronecker delta, and $s$, $\alpha$ and $\tau$ are the ``hyperparameters'' of the noise model (i.e. parameters we wish to marginalize over).

In this framework, every likelihood computation requires an evaluation of both $K^{-1}$ and $\det\,K$. Na\"ievely, the computational complexity of this operation scales as $n^4$, where $n$ is the number of data points. This computation is generally intractable for datasets like \kep light curves, especially short cadence observations. We exploit recent developments in the applied math literature \citep{2014A} to achieve $\mathcal{O}(n^2\,\log n)$ computational scaling on both short and long cadence datasets (Foreman-Mackey et al.\ in prep). This method is analogous to the technique proposed by \citet{2012G} but these algorithmic advancements allow posterior sampling and marginalization on datasets larger than previously possible.

In practice, we assume that the long cadence datasets share one set of hyperparameters $(s_\mathrm{LC},\,\alpha_\mathrm{LC},\,\tau_\mathrm{LC})$ and that the short cadence datasets are described by an independent parameter set $(s_\mathrm{SC},\,\alpha_\mathrm{SC},\,\tau_\mathrm{SC})$. These six hyperparameters are included as dimensions in our MCMC sampling so the results have properly taken uncertainties in the noise model into account. For this model, we draw MCMC samples from the posterior probability density using the {\tt emcee} package \citep{2013FH}, and the results are reported in Tables~\ref{tab:lc}, \ref{tab:lcttv}, and \ref{tab:lcb}.

\begin{figure}
\includegraphics[width=.5\textwidth]{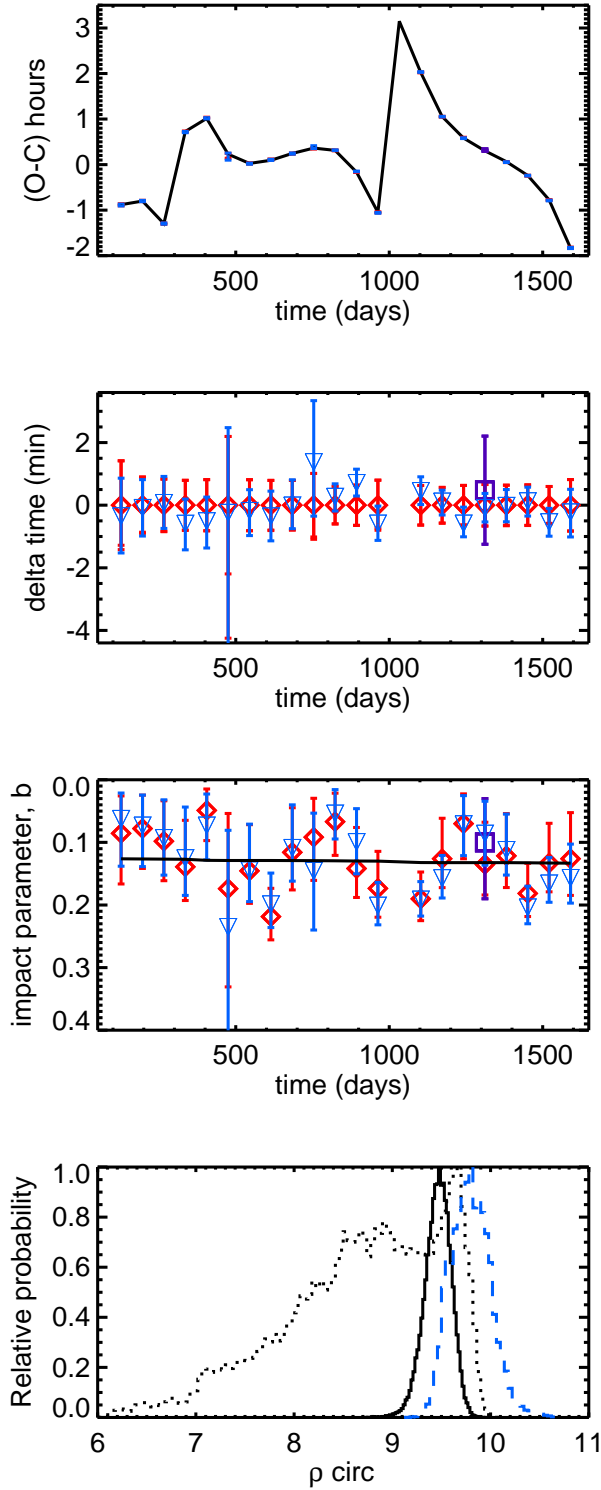}
\caption{\footnotesize  Top panel: TTVs with best-fitting model (Table \ref{tab:bestfit}). Second panel: difference between the TTVs derived from our first fitting approach and our second. Red points come from our first fitting approach and blue from our second (Table \ref{tab:lc}, \ref{tab:lcttv}, \ref{tab:lcb}). The purple squares are from ground-based light curves (Section \ref{sec:ground}). Third panel: impact parameters. Bottom panel: $\rhocirc$ posterior, allowing the impact parameter of each transit to vary (solid) and forcing the impact parameters to be the same (dotted). This panel illustrates the necessity of fitting a different impact parameter to each light curve; otherwise the $\rhocirc$ posterior becomes very wide to try to compensate for the different transit durations.\label{fig:ttdur}}
\end{figure}

In Figure \ref{fig:ttdur}, we plot the TTVs, the impact parameters, and the posteriors for $\rhocirc$. The transit with a large error bar is a partial transit cut off by a gap in the observations. The TTVs are large and reveal a period of $\sim700$ day for the perturbing companion, whose properties we will precisely measure in Sections 4 and 5. The impact parameters deviate only subtly from a constant value. However, if we require the transiting planet's impact parameter to be constant, the posterior for $\rhocirc$ becomes very wide to compensate for the different transition durations. { In the bottom panel of Figure \ref{fig:ttdur}, we plot the posterior for $\rhocirc$ in our nominal fit, in which we allow the impact parameter of each transit to vary (solid line; Tables~\ref{tab:lc}  and \ref{tab:lcb}), and from an alternative fit in which we force $b$ to be the same for each light curve (dotted line). The latter posterior for $\rhocirc$ is much wider and less smooth.}

\begin{figure}
\includegraphics{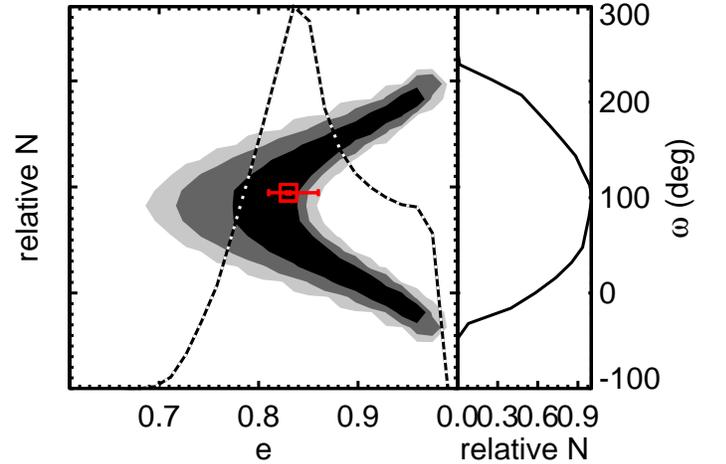}
\caption{Left: Joint posterior for $\omega$ vs. $e$ from the photoeccentric effect (Table \ref{tab:lc}). The black (gray, light gray) contours represent the $\{68.3,95,99\}$\% probability density levels (i.e. 68$\%$ of the posterior is contained within the black contour). Over-plotted as a black-and-white dotted line is a histogram of the eccentricity posterior probability distribution marginalized over $\omega$. The red square marks the best-fitting value with error bars that we measure independently from the RVs { in Section \ref{sec:threed}, Table \ref{tab:bestfit}}. The eccentricity measured from the photoeccentric effect is in good agreement with that from the RVs. Right: Posterior distribution for $\omega$, marginalized over eccentricity.  \label{fig:eccom} }
\end{figure}

To derive the eccentricity posterior, we combine the posterior for $\rhocirc$ derived from the transit light curves with the posterior for the true stellar density from Section \ref{sec:sme}, following the procedure
\footnote{\citet{2014K} derives a conservative criterion (Equation B11 of \citealt {2014K}) for which certain small-angle approximations (assumed by this procedure) hold. Although Kepler-419b technically in violation of that criterion, in \citet{2012DMJ} Appendix F, we recast that criterion in terms of the $g$ measured from a circular fit, finding that the approximations are appropriate for $g = \frac{1+e\sin\omega}{\sqrt{1-e^2}} =  (\frac{\rhocirc}{\rho_\star})^{1/3}< 19$. Since we derive $g=3$ (Table \ref{tab:lc}), the approximation is appropriate for Kepler-419b. In Section \ref{sec:threed}, we will confirm this eccentricity using RV measurements.} \citet{2012DJ}, Section 3.4. From the light curves, we measure the planet's eccentricity to be e =  0.85$^{+0.08}_{-0.07}$, consistent with value of $e = 0.81^{+0.10}_{-0.07}$ reported by D12. The change in the value and its uncertainties are mostly due to the improved $\rho_\star$ (Section \ref{sec:star}) but partly due to the tighter constrain on $\rhocirc$ that we derived when we { allowed} the impact parameter to vary among the light curves. In Figure \ref{fig:eccom}, we plot the eccentricity and $\omega$ posteriors, marking the value  that we will measure independently from the RVs (Section \ref{sec:threed}; Table \ref{tab:bestfit}), which is in good agreement.

\begin{deluxetable}{rrlrl}
\tabletypesize{\footnotesize}%
\tablecaption{Planet Parameters for KOI 1474b Derived from the Light-curves \label{tab:lc}}
\tablewidth{0pt}
\tablehead{
\colhead{Parameter}    & \colhead{Value\tablenotemark{a}}}
\startdata
												& Median-filter/CW09/{\tt TAP}\tablenotemark{b}		&				& Gaussian processes\tablenotemark{c}/{\tt emcee}\tablenotemark{d} \\
\hline
\\
Planet-to-star radius ratio, $R_{p}/R_{\star}$   				&0.0626 &$\pm$ 0.0002 		&0.06187&$^{+0.00016}_{-0.00018}$	\\\\
Light curves stellar density, $\rhocirc$  [$\rho_\odot$]  					&9.47 &$^{+0.13}_{-0.14}$ 	&9.8 &$\pm 0.2$ 			 	\\\\
Density ratio parameter, $g=(\frac{\rhocirc}{\rho_\star})^{1/3}$ 	&3.0 &$\pm0.3$			& 3.0	&$\pm0.3$\\\\
Limb darkening coefficient, $q_{1}$ 						& $0.37$& $^{+0.08}_{-0.07}$ 	&0.6&$^{+0.3}_{-0.2}$				\\\\
Limb darkening coefficient, $q_{2}$ 						& $0.24$&$^{+0.07}_{-0.06}$	& $0.21$&$\pm0.04$			\\\\
Planetary radius, $R_{p}$ [$R_{\oplus}$] 					&10.8 &$^{+1.4}_{-1.3}$ 		&10.7 &$^{+1.5}_{-1.4}$ \\\\
Normalized red noise, short-cadence, $\sigma_r$	[ppm]		& 4530 & $\pm180$  		\\\\
Normalized white noise, short-cadence $\sigma_w$	[ppm]		& 651&$\pm 3$  		\\\\
Normalized red noise, long-cadence $\sigma_r$		[ppm]	& 400&$\pm 5$  		\\\\
Normalized white noise, long-cadence $\sigma_w$	[ppm]		& 128&$\pm 4$  		\\\\
Extra white noise, short-cadence $s_\mathrm{SC}$ [ppm]			&&							& 142&$\pm11$ \\\\
Lag, short-cadence,$\tau_\mathrm{SC}$		[days]		&&							& 0.33&$^{+0.03}_{-0.02}$\\\\
Red noise amplitude, short-cadence $\alpha_\mathrm{SC}$[ppm]	&&							&310&$\pm 30$\\\\
Extra white noise, long-cadence $s_\mathrm{LC}$	[ppm]		&&							& 69 & $\pm4$ \\\\
Lag, long-cadence,$\tau_\mathrm{LC}$		[days]			&&							& 0.34&$\pm 0.02$\\\\
Red noise amplitude, long-cadence $\alpha_\mathrm{LC}$[ppm]	&&							&260&$\pm 20$\\\\
Eccentricity, $e$ 									& 0.85&$^{+0.08}_{-0.07}$ 	&	0.86&$^{+0.08}_{-0.06}$														\enddata
\tablenotetext{a}{The uncertainties represent the 68.3\% confidence interval of the posterior distribution.}
\tablenotetext{b}{{\tt TAP} software by \citet{2011G}. Uses CW09 wavelet likelihood.}
\tablenotetext{c}{Daniel Foreman-Mackey et al., in prep}
\tablenotetext{d}{\citet{2013FH}}
\end{deluxetable}
\clearpage
\section{Ground-based follow-up}
\label{sec:ground}

{ Although data acquisition on the original \kep field by the \kep spacecraft has ended}, there are a number of systems for which additional TTVs would greatly improve the precision of the derived mass measurement and/or clarify the qualitative picture of the dynamics. With its large TTVs and host star brighter than most \kep host stars (\kep magnitude 13.0), Kepler-419b is a case study for whether ground-based follow-up can allow for sufficient precision. Here we observe transits from the ground at the same time as a \kep transit and compare the transit time and impact parameter we measure. We observe using four telescopes: the Nickel 1-meter telescope at Lick Observatory in Mountain Hamilton, CA and three telescopes that are part of the Las Cumbres Observatory Global Telescope Network (LCOGT; \citealt{2013BBB}): the Faulkes Telescope North (FTN) 2-meter at Haleakala Observatory in Hawaii, the Byrne Observatory at Sedgwick (BOS) 0.8-meter at the Sedgwick Reserve in the Santa Ynez Valley, CA, and the El Paso (ELP) 1-meter at { McDonald} Observatory in Fort Davis, Texas. The observations were all taken on the night of August 4th-5th, 2012.

The observations at the Nickel telescope yielded the highest precision light curve but, due to the timing of sunset, only a partial transit. The Nickel exposures were taken using the Direct Imaging Camera with 2x2 pixel binning, fast readout mode, and the \citet{1990B} I filter. The telescope was defocused to achieve a mountain-shaped point spread function, maximizing the number of pixels across the PSF while optimizing the exposure time to achieve high cadence yet keep the readout time a modest fraction of the exposure time. The focus was kept at a constant focus position of 367 throughout the night. The sky was overcast at sunset but cleared by midnight, and sky flats were taken at dawn. Images were taken continuously with an exposure time of 180 s and 5 s read-out time, yielding an out-of-transit scatter of 830 ppm and light curve model residuals of 843 ppm, corresponding to 1.5 mmag/minute.

The LCOGT Network observations were taken continuously in the SDSS r' filter with a 40 s exposure time with the FTN and 120 s exposure time with the BOS and ELP, yielding out-of-transit scatter of 2703 ppm, 2082 ppm, and 2641 ppm and model residuals of 2695 ppm, 2062 ppm, and 2205 ppm at FTN, BOS, and ELP respectively. For comparison, the photometric noise rates are 2600, 3300, and 4500 ppm per minute for the FTN, BOS, and ELP respectively \citep{2011F}.  Of the four sets of ground-based observations, only those at ELP cover the entire transit (due to the eastern longitude of the telescope); at this telescope, the target drifted over 2 pixels throughout the observation, causing correlated noise originating from non-perfect flat field correction. { We did not apply corrections for the star's position for any of the light curves presented here.}

\begin{figure}
\includegraphics{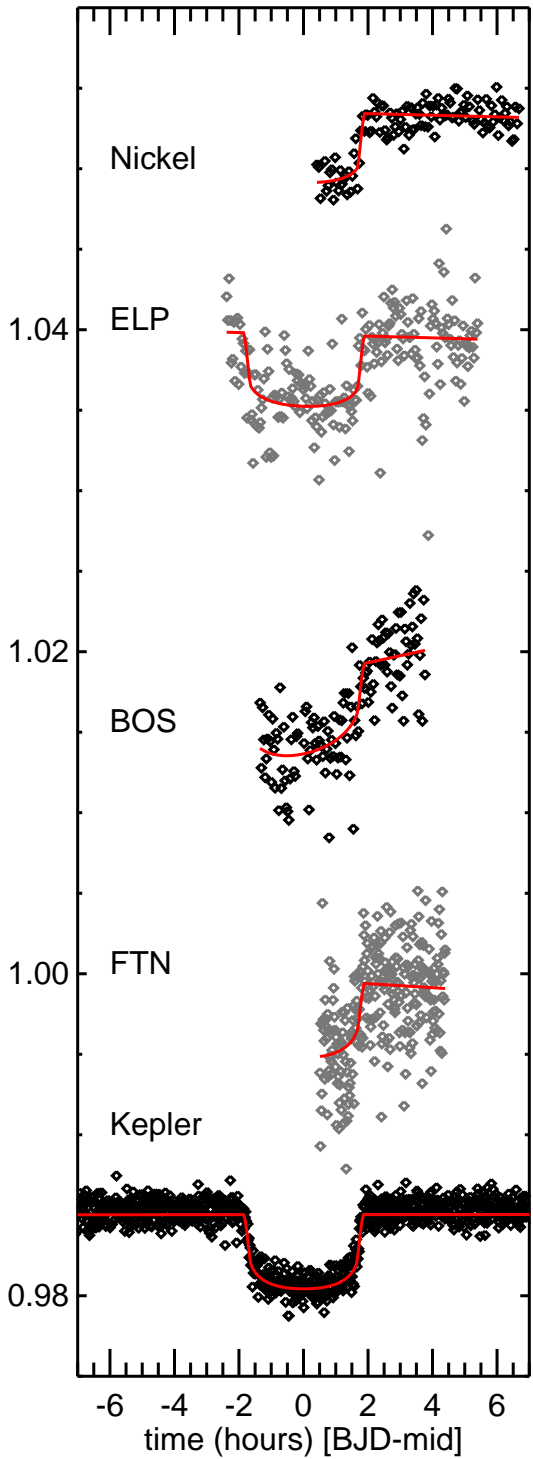}
\caption{Diamonds: Light curves (relative flux) observed by Nickel (top), ELP (second), BOS (third), FTN (fourth), simultaneous \kep (bottom), offset vertically for clarity. Red line: best-fitting model (Table \ref{tab:fits}). \label{fig:groundlc}}
\end{figure}

{ We fit the ground-based light curves simultaneously with the \kep light curves, forcing four ground-based light curves to share a common transit time and impact parameter but allowing the quantities to differ from the simultaneous \kep light curve (mimicking a future situation in which have ground-based light curves without simultaneous \kep photometry). }We allow each ground-based light curve to have its own red and white noise parameter and linear trend. In Figure \ref{fig:groundlc}, we plot the light curves and best-fitting models. The ground-based mid-transit time is 1311.7275$\pm$0.0012 days [BJD-2454833], in good agreement with the \kep light curve value of $1311.7272\pm0.0005$. The mid-transit time is precise to 1.8 minutes, sufficient to measure the hour amplitude of the TTVs of Kepler-419b to high precision. We measure an impact parameter of of $0.10^{0.09}_{-0.07}$, in agreement with the measurement of $0.26^{0.11}_{-0.15}$ from the \kep light curves. In Figure \ref{fig:ttdur}, we overplot the measurements from the ground as purple squares. Given the good agreement and reasonable precision of the ground-based transit time and duration, we consider Kepler-419 to be a case study demonstrating that ground-based follow-up is feasible for high S/N planetary transits with large TTVs, sufficiently well-predicted transit times that we know on which night to observe, and short enough transit durations to be covered in a night.

\section{Confirmation of inner planet's existence and eccentricity by RV}
\label{sec:RVs}

\begin{figure*}
\includegraphics{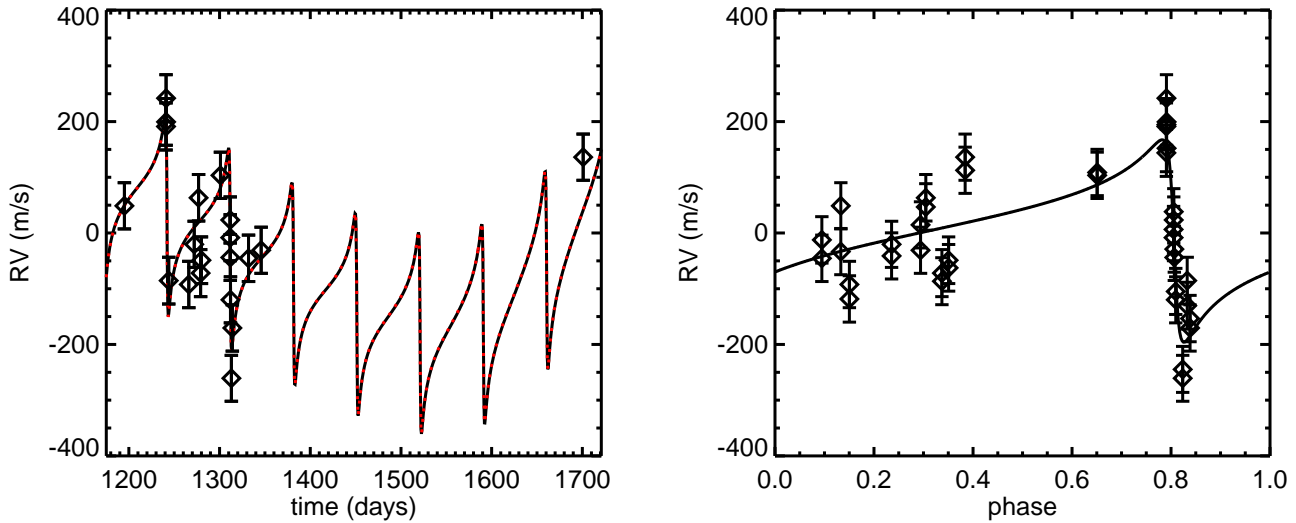}
\caption{Left: RV measurements (diamonds, Table \ref{tab:RVs}); best-fitting model from joint fit to TTVs/b/RVs (solid black; Table \ref{tab:bestfit}); best-fitting coplanar model to the TTVs alone (red dotted; Table \ref{tab:fits}, column 1). We add 40 m/s of stellar jitter in quadrature to the error bars. Note: The TTVs do not constrain the systematic velocity or the inner planet's mass; we set these to the best-fitting values from the joint fit. For ease of comparison, we also set the eccentricity and periapse of the inner planet to the best-fitting value from the joint fit, but this value is at the center of the two-dimensional posterior estimated from the photoeccentric effect (Figure \ref{fig:eccom}). Right: RVs phased to the inner planet's orbital period, with the trend from the outer-planet subtracted off.\label{fig:RVs}}
\end{figure*}

{ Previously (D12), we reported a 3.1\% false positive probability for Kepler-419b, based on the validation procedure developed by \citet{2012M}. The most likely false positive scenario was a hierarchical eclipsing binary. Here we confirm Kepler-419b's planetary nature and eccentricity with RV measurements. We obtain spectroscopic observations at Keck observatory using HIRES \citep{1994V}. The spectra were observed using the standard setup of the California Planet Survey \citep{2010HJ,2010JH}. The observations span May 2012 through August 2013, with the majority taken during summer 2012. Exposure times range from 800 to 1200 seconds, depending on the weather and seeing. } We { list the RV measurements in Table \ref{tab:RVs} and plot them} in Figure \ref{fig:RVs}. The uncertainties have 40 m/s stellar jitter added in quadrature to account for the additional scatter in this noisy F star. The 40 m/s is estimated from the scatter itself. We are developing a noise model for future work that better accounts for the correlated stellar noise.

\begin{deluxetable}{llll}
\tabletypesize{\footnotesize}%
\tablecaption{Keck HIRES RV Measurements of Kepler-419\label{tab:RVs}}
\tablewidth{0pt}
\tablehead{
\colhead{Time [BJD-2454833]}  & \colhead{Value (m/s)}& \colhead{Uncertainty (m/s)} & { S/N}}
\startdata
\hline
\\
       1195.1136&       48.8&       10.9 & 70\\
       1240.9869&       191.2&       13.5 & 45\\
       1240.9976&       241.9&       13.9 & 47\\
       1241.0088&       199.4&       12.8 & 48\\
       1243.9177&      -85.5&       12.8 & 52\\
       1266.0687&      -92.4&       11.5 & 58\\
       1272.0193&      -20.3&       10.7 & 58\\
       1276.8389&       63.2&       11.9 & 58\\
       1279.1070&      -72.0&       13.4 & 55\\
       1280.0553&      -48.9&       12.5 & 77\\
       1300.9847&       103.5&       11.3 & 72\\
       1311.7530&      -8.4&       10.3 & 72\\
       1311.7800&       23.4&       10.8 & 72\\
       1311.8565&      -43.9&       10.2 & 71\\
       1312.0579&      -119.9&       10.4 & 70\\
       1313.0389&      -260.8&       10.2 & 70\\
       1314.0708&      -170.4&       11.3 &58 \\
       1331.9493&      -45.4&       11.7 & 58\\
       1345.8502&      -30.9&       10.7 & 72\\
       1700.9169&      136.1&      11.0 & 72\\	
\enddata
\end{deluxetable}

The measurements reveal three important features: 1) an amplitude of a couple hundred m/s, corresponding to several Jupiter-mass planet at a fraction of an AU, confirming the planetary nature of Kepler-419b, 2) near the transit time at 1311 days, when we took a high density of measurements, the RV decreases rapidly, { indicating the star's reflex motion to a highly eccentric planet undergoing periapse passage} with the periapse pointed toward us: over a short interval, the planet changes from moving toward us to moving parallel to moving away from us, and 3) a trend consistent with the presence of a longer-period outer companion, presumably the non-transiting planet we detected and characterized through TTVs. Based on the first and second features, Kepler-419b is transiting near periapse and has an eccentricity consistent with that derived from the ``photoeccentric" effect (Figure \ref{fig:eccom}). Therefore the RV measurements confirm Kepler-419b and its high eccentricity and are consistent the presence of the non-transiting planet, Kepler-419c, detected via Kepler-419b's TTVs. For example, a 21 Jupiter mass brown-dwarf perturber with the same orbit { (see Table \ref{tab:bestfit} for orbit)} would produce a 1 km/s RV variation over the observed timescale, inconsistent with the RV observations. However, we distinguish that the RVs do not \emph{independently} confirm either planet; without constraints on the periods and epochs from the transit, the RVs currently lack the coverage and precision for us to derive the complete set of properties of the two planets from the RVs alone. The RV measurements also put limits on additional perturbers; based on the observed acceleration, we can rule out 0.2 solar mass companion closer than 10 AU (e.g. \citealt{2008Q}, Equation 5).

\section{Three-dimensional architecture from dynamical fits}
\label{sec:threed}
{ Here we simultaneously fit the transit times, impact parameters, radial velocities, and $\rhocirc$ to obtain three-dimensional orbits for both planets. The constraints on the outer planet's mass and orbit come primarily from the TTVs, with the impact parameter and RVs adding no additional constraints. In practice, the impact parameters only constrain the inner planet's inclination relative to our line of sight. The RVs allow us to measure the inner planet's mass (to which the TTVs are not at all sensitive) and confirm and more tightly constrain the inner planet's eccentricity and periapse originally derived from the photoeccentric effect (Section \ref{sec:lc}). The RVs do not contribute much to our knowledge of the outer planet because its mass and three-dimensional orbit are already tightly constrained by the TTVs, and the RVs have not yet covered a full orbital period of the outer planet. }

The reason that is possible to uniquely determine the mass and three dimensional orbit of the outer planet from the TTVs is that the variation in the transit times of the inner planet (Fig. \ref{fig:ttdur}, top panel) are a tidal effect caused by the varying position of the outer planet. The effective radial force (which can be conceptualized as the effective stellar mass) felt by the inner planet (Kepler-419b) changes with the position of the outer planet (Kepler-419c) because of three effects. First, Kepler-419b spends most of its time at apoapse, and, moreover, the force between Kepler-419b and Kepler-419c at their conjunction is strongest if Kepler-419b is at apoapse. Therefore the effective radial force on Kepler-419b varies depending on angular separation of Kepler-419c from Kepler-419b's apoapse. This effect would cause TTVs even if planet c were on a perfectly co-planar, circular orbit. Second, Kepler-419c's eccentricity causes variations in its separation from Kepler-419b, as Kepler-419c moves from its periapse to its apoapse. Finally, because planet c is not perfectly coplanar, its force projected onto Kepler-419b's position vector changes as the planet c moves above and below the orbital plane of Kepler-419b; any mutual inclination would cause TTVs even if both planets were on circular orbits. In Appendix \ref{app:ttvs}, we illustrate these effects and justify why we can measure the perturber's properties, including its mutual inclination, solely from the TTVs without degeneracies.

{ To fit the TTVs, we integrate the gravitational forces} using the Position Extended Forest-Ruth Like (PEFRL) algorithm \citep{2002O}, a fourth order symplectic integrator. We choose this integrator because it is both symplectic and high order, allowing us to accurately compute tens or hundreds of millions of models for our fitting process on a reasonable computing timescale of a few days. Because the inner planet's orbit is highly eccentric, it is inefficient to sample the entire orbit with the tiny time steps required to resolve the periapse passage. Therefore we follow \citealt{1999R} (1999, Section 3.1; see also \citealt{1997M}) and sample in $ds = dt /r$ instead of $dt$, where $ds$ is a time-regularized step, $dt$ is a step in time and $r$ is the separation between Kepler-419b and its host star. Then we find the precise transit times using the iterative algorithm described in Section 2.5 of \citet{2010F}.

{ We perform a joint MCMC fit to the TTVs, RVs, $b$, and $\rhocirc$. { To demonstrate which data are constraining which orbital properties, we perform fits to subsets of these data in Appendix \ref{app:ttvfits}.} We report the parameters in Table \ref{tab:bestfit} and including the mass $m$, eccentricity $e$, argument periapse in the sky $\omega$, inclination relative to the line of sight $i$, longitude of ascending node in the sky plane $\Omega$, and mean anomaly $M$. See Appendix \ref{app:ttvs} for a diagram of orbital elements. All orbital elements are osculating (epoch BJD 2455809.4009671761741629) and Jacobian. All priors are uniform, here and throughout the paper, unless otherwise specified; the uniform priors on angles are equivalent to an isotropic prior on the orientation of each planet's orbit. The derived values are similar to coplanar fit in D12 based on eight quarters of data, but with the additional TTVs, we can constrain Kepler-419c's mass and mutual inclination. The perturbing companion, Kepler-419c, is a moderately eccentric ($e_c =  0.184^{+0.002}_{-0.002}$) giant planet ($m_c = 7.3\pm0.4$ Jupiter masses), located at $1.68\pm0.03$ AU. The \kep data rule out transits of Kepler-419c at its conjunction epoch { at 1327.14 + $n$675.46 days [BJD-2454833], where $n$ is an integer and we exclude solutions for which Kepler-419c would transit from the posterior.}} { We constrain the outer planet's inclination relative to the line sight to be $i_c = 88^{+3}_{-2}$ degrees and its longitude of ascending node relative to the sky plane to be $\Omega_c = 4^{+12}_{-12}$ degrees, corresponding to a low mutual inclination relative to Kepler-419b of $i_{\rm mut} = 9^{+8}_{-6}$ degrees, with a 99\% confidence upper limit of 27 degrees. (See Appendix \ref{app:ttvfits} and \ref{app:ttvs} for a detailed exploration of how these quantities are constrained by the TTV signal.) We perform an identical fit except using the SPC stellar parameters (Appendix \ref{app:spec}) and find the values are consistent (Table \ref{tab:fits}, right column).} 

{ In addition to the data, we favor of the coplanar solution for two other reasons. First, despite the fact that the second planet does not transit, its inclination relative to the line of sight ($i_c$) is very close to that of the transiting planet ($i_b$), independent of $\Omega_c$. This would be a surprising, fine-tuned coincidence if the planets were non-coplanar. Second, the difference between the arguments of periapse in the sky-plane is very close to perfectly anti-aligned\footnote{The libration amplitude is highly sensitive to the uncertainties in $e_b$ and $\omega_b$ but typically $< 50^\circ$.} ($179^\circ.8^{+0.6}_{-0.6}$). For the geometry here, if the system were coplanar, this would correspond to separation of periapses in the invariable plane, a quantity that librates about 180 degrees in many parts of parameter space, e.g. \citet{2004M}. If the system were non-coplanar, it would be a strange coincidence. A similar argument was made for the coplanarity of Upsilon Andromeda by \citet{2001C}.}

\begin{deluxetable}{rrr}

\tabletypesize{\small}%
\tablecaption{Planet Parameters for Kepler-419b and Kepler-419-c at epoch BJD 2455809.4009671761741629. All orbital elements are Jacobian.\label{tab:bestfit}}
\tablewidth{0pt}
\tablehead{
\colhead{Parameter}    				& \colhead{Value}}
\startdata
$m_\star (m_\odot)\tablenotemark{a}	$ 	&1.39&$^{+0.08}_{-0.07}$ \\
$R_\star \tablenotemark{a}$			&1.75&$^{+0.08}_{-0.07}$	\\
$m_{b} (M_{\rm Jup})$				&  2.5 &$\pm$ 0.3			\\
$P_{b}$ (days)					&  69.7546&$^{+0.0007}_{-0.0009}$\\
$a_{b}$ (AU) \tablenotemark{b}		&  0.370&$^{+0.007}_{-0.006}$	\\
$e_{b}$						&  0.833&$^{+0.013}_{-0.013}$	\\
$\omega_{b}(^\circ)$				&  95.2&$^{+1.0}_{-1.2}$		\\
$M_{b}(^\circ)$					&  68.69&$^{+0.05}_{-0.05}	$	\\
$i_{b} (^\circ)$					&  88.95&$^{+0.14}_{-0.17}$		\\		
$\Omega_{b}(^\circ)$			&   0 &(fixed)\\
$m_{c} (M_{\rm Jup})$			&  7.3&$\pm0.4$		\\
$P_{c} (days)$					& 675.47&$^{+0.11}_{-0.11}$	\\
$a_{c}$ (AU) 		\tablenotemark{b} & 1.68&$\pm$0.03	\\
$e_{c}$						& 0.184&$^{+0.002}_{-0.002}$	\\
$\omega_{c} (^\circ)$			& 275.3&$^{+1.2}_{-1.0}$	\\
$M_{c} (^\circ)$					& 345.0&$\pm0.3$	\\
$\Omega_{c} (^\circ)$			&4&$^{+12}_{-12}$	\\
$i_{c} (^\circ)$\					&88&$^{+3}_{-2}$	\\		
$i_{\rm mut} (^\circ)$				&9&$^{+8}_{-6}$	\\
$99\%$ $i_{\rm mut} (^\circ)$		&27			\\
$\omega_{b}- \omega_{c} (^\circ)$	&179.8&$^{+0.6}_{-0.6}$\\
$\varpi_{b}- \varpi_{c} (^\circ)$		&176&$^{+12}_{-12}$	\\
Systemic offset (m/s)				& -32&$\pm 10$	\\
\enddata
\tablenotetext{a} {Posterior from Section \ref{sec:star} imposed as a prior.}
\tablenotetext{b} {Derived from stellar mass and orbital period posteriors.}
\end{deluxetable}

{ In Appendix \ref{app:is}, we constrain the inner planet's spin-orbit alignment from projected rotational velocity, finding some evidence that the entire system is misaligned with the host star's spin axis. However, better modeling of the radial-velocity noise is necessary to confirm this conclusion.}

\section{Migration scenarios}
\label{sec:mig}
With a small semi-major axis $(a=0.370$ AU) interior { to the observed} pile-up of giant planets at 1 AU \citep{2008C} and { to} the several AU beyond which giant planets are thought to form (e.g. \citealt{2008KB}) and { with} a periapse too distant for tidal circularization, Kepler-419b is a member of the ``Period Valley" population whose dynamical origin is mysterious (e.g. \citealt{2013D}) . Here we consider several scenarios for the origin of Period Valley planets and whether they are consistent with the properties of the Kepler-419 system derived here.

\begin{figure*}
\includegraphics{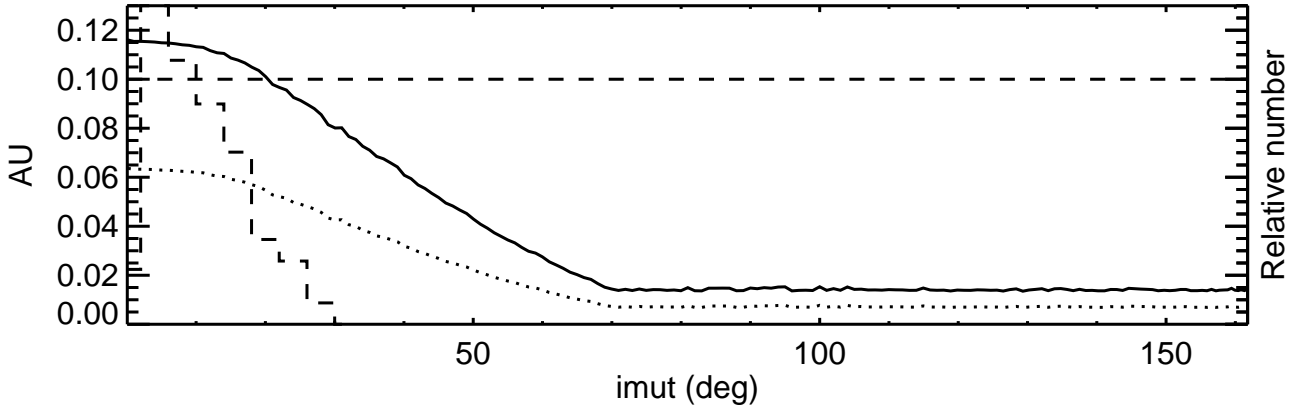}
\caption{Minimum $\afinal = a (1-e^2)$ (solid) and periapse $a(1-e)$ (dotted) as a function of mutual inclination (imut) from the long term integrations described in Appendix \ref{app:stable}, corresponding to Figure \ref{fig:imut} and using fits to the TTVs/b/RVs/$\rhocirc$ .The dashed line is the posterior distribution of the mutual inclination between planet b and c from the fit in Table \ref{tab:bestfit}; the right y-axis refers to the posterior. The horizontal dashed line represents the cut-off employed by \citet{2012SKDT}, \citet{2012DMJ}, and \citet{2014D} for planets that may have periapses close enough to tidally circularize.  The minima asymptote to the stellar radius because the planet is assumed to not survive a collision with the star. \label{fig:minaf}}
\end{figure*}

\citet{2014D} suggested that the Period Valley planets are undergoing Kozai cycles and periodically reach eccentricities high enough to migrate via tidal circularization. Under this theory, we are currently observing the Period Valley planets in the low-eccentricity phase of their cycle. With our previous dataset (D12), this was a possibility for Kepler-419b because we did not know the mutual inclination of Kepler-419c; moreover, Kepler-419b was a particularly promising candidate for this scenario because its observed eccentricity is large and only needs a small boast to reach $\afinal = a (1-e^2)$ 0.1 AU, the maximum $\afinal$ for tidal circularization over a typical host star lifetime employed by \citet{2012SKDT}, \citet{2012DMJ}, and \citet{2014D}. However, this dynamical evolution is inconsistent with the low mutual inclination that we have now measured for Kepler-419b and c (Sections \ref{sec:threed}). In Figure \ref{fig:minaf}, we plot the minimum periapse  achieved by Kepler-419b --- over the course many secular oscillations --- as a function of the mutual inclination with planet c. In order for Kepler-419b to be a proto-hot-Jupiter that gets sufficiently close to its star to tidally circularize but not collide with its star, it would need to periodically reach a minimum $ \afinal < 0.1$ AU. Below a mutual inclination 21$^\circ$, Kepler-419b does not get close enough to the star to tidally circularize. Above $70^\circ$, the planet collides with the star. A mutual inclination between 21 and 70 degrees is not a good fit { to} the data (Table \ref{tab:bestfit}, Figure \ref{fig:imut}, Table \ref{tab:fits}). Consequently, we conclude that oscillations in Kepler-419b's eccentricity due to Kepler-419c are not causing Kepler-419b's orbit to shrink and circularize.

There are several qualifications to this conclusion. First, we can only rule { out} a mutual inclination above 21 degrees at the { 91\%} confidence level (Table \ref{tab:bestfit}), so there is a small chance that Kepler-419b can reach $\afinal < 0.1$ AU. The uncertainty in the mutual inclination is attributable to the uncertainty in Kepler-419b's eccentricity (i.e. this high mutual inclination is ruled out at 99.8\% confidence level with the inner planet's eccentricity fixed at the best fit value, Table \ref{tab:fits}, column 2), so additional RV measurements and better modeling of the stellar noise in the RVs should allow us to distinguish in the future. Second, the presence of a { fourth body}, an undiscovered planet { or star}, could affect the dynamics. A  { fourth body} in the system could cause additional oscillations in Kepler-419b's eccentricity, allowing it to reach a higher value. { The fourth body would need to be massive and nearby enough} for its secular mode to contribute significantly to Kepler-419b's eccentricity. { A nearby planet would have be undetected in the TTVs and RVs, and a stellar binary companion, e.g. \citet{2008T}, would have be undetected by our adaptive-optics observations (D12) and a second set of lines in the stellar spectra. We plan to exhaustively explore the parameter space in future work.} Finally, Kepler-419c could have caused high-eccentricity migration of Kepler-419b in the past if Kepler-419c used to have a larger eccentricity, smaller semi-major axis, or larger mutual inclination but then had its orbit altered by a third planet.

\citet{2014L} describe a mechanism by which the inner, less massive planet in an initially coplanar but eccentric system can have its mutual inclination flipped and, in the process, undergo close passages to the host star and tidally circularize. Although the Kepler-419 system meets the analytical criterion for triggering this mechanism derived by \citet{2014L}, we do not observe such flips occurring in our long-term integrations and therefore rule out this possibility. We expect that the approximations that the system is hierarchical and that the inner planet is a test particle do not apply to this system, whose planets have a semi-major axis ratio of about 4 and mass ratio of about 3. See \citet{2013T} for an exploration of the parameter space for flips in the large mutual inclination regime.

Several possibilities remain that could account for Kepler-419b's small semi-major axis and large eccentricity while remaining consistent with the low inclination with Kepler-419c. One is disk migration followed by planet-planet scattering (e.g. \citet{2011GR}). Although \citet{2014P} demonstrate that warm Jupiters with $a< 0.15$ AU could not have their eccentricities excited to the observed level at their present-day semi-major axes because their escape velocities are much smaller than their circular velocities, Kepler-419b is not in this regime. Dynamical instability in the presence of a gas disk \citep{2013L}, possibly triggered by resonance crossings, could cause Kepler-419b to migrate in yet achieve an eccentric orbit. This scenario is attractive because it could account for the apsidal anti-alignment of Kepler-419b and Kepler-419c, which are anti-aligned to within $0.2 \pm 0.6$ degrees in the coplanar case { (see discussion at the end of Section \ref{sec:threed})}. This tight anti-alignment is unlikely to be the result of observational bias (i.e. we could still constrain the outer planet's identity if it were not apsidally anti-aligned; see Appendix \ref{app:ttvs}). Disk dissipation can drive systems to apsidal alignment or anti-alignment \citep{2002C,2013Z}. Another possible scenario is that torques from the gas disk led to growth in Kepler-419b's eccentricity \citep{2003G,2004S}. \citet{2013DA} found in high resolution, three-dimensional simulations that such growth does not occur for planets of Kepler-419b's mass embedded in isoentropic disks, but \citet{2014T} recently showed that eccentricity growth could occur if the disk is non-isoentropic due to shadowing. However, it has yet to be demonstrated that such growth could lead to an eccentricity as large as 0.8. Without the involvement of a gas disk, planet-planet scattering would require several to tens of planets of Kepler-419b's own mass to be ejected or scattered out to large semi-major axes. Secular chaos would likely also require additional planets to achieve Kepler-419b's large angular momentum deficit if the planets initially had low $(e<0.1)$ eccentricity. Both planet-planet scattering and secular chaos produce a range of mutual inclinations \citep{2008CF,2011WL}, and the Kepler-419 planets would have to coincidentally be at the low end of the range.

In summary, Kepler-419b is not undergoing high eccentricity migration (in which a planet evolves from a large semi-major and large eccentricity to a close-in, circular orbit through tidal circularization), even at the minimum periapse it reaches over the course of its secular evolution. Planet-planet scattering or secular chaos remain possibilities for explaining Kepler-419b's large eccentricity and small semi-major axis but it is necessary to invoke a gas disk or additional planets.

\section{Summary and future work}
\label{sec:summary}
Using information from transit light curves---including transit timing variations and the photoeccentric effect---we mapped out the three-dimensional architecture of the Kepler-419 system, which hosts two giant planets. The transiting, inner, $2.5 \pm 0.3$ Jupiter-mass planet's orbit has a large eccentricity $(e=0.824^{+0.019}_{-0.010}$) and small semi-major axis (0.362$^{+0.006}_{-0.007}$ AU); the non-transiting outer planet is more massive (7.3$\pm$0.4 Jupiter masses) and is hierarchically separated (1.68$\pm$0.03 AU). RV measurements allowed us to confirm the inner planet's large eccentricity and are consistent with the presence of the outer planet. Surprisingly, the planets in this eccentric, hierarchical system are close to coplanar (mutual inclination { $9^{+8}_{-6}$ } degrees). The inner planet's close-in orbit and large eccentricity are most likely not a product solely of the processes of high-eccentricity migration (tidal friction shrinking and circularizing an initially large eccentricity, large semi-major axis orbit), even accounting for the planet's secular evolution, including eccentricity oscillations caused by Kepler-419c. It remains a possibility that there is a { fourth body} in the system causing oscillations in Kepler-419b's eccentricity yet not currently detected in the TTVs, RVs, { adaptive optic images (D12), or stellar spectrum}; such a body would need to be massive and nearby enough to contribute precession comparable to that from Kepler-419c. We recommend continued radial velocity-follow up to place better constraints on the presence of a massive companion with an orbital period of several years or more. Moreover, in Section \ref{sec:mig}, we concluded that if Kepler-419b achieved its high eccentricity orbit through planet-planet scattering or secular chaos, other planets (besides Kepler-419c) and/or a gas disk were most likely involved. 

Although Kepler-419 itself is just one data point, if other eccentric, hierarchical systems are found to have low mutual inclinations, this may call into question the interpretation that the extreme spin-orbit misalignments observed for hot Jupiters are the result of the planet's orbit being tilted out of the plane it formed in by scattering, Kozai, or secular chaos. Additional theoretical work is needed to simulate dynamical scenarios for producing Kepler-419b and other ``Period Valley'' $0.1 < a < 1$ giant planets, including planet scattering and secular chaos, with or without the presence of a gas disk. Our result is thematically related to \citet{2012TD}'s finding that the \kep systems are consistent with being drawn from the (often quite eccentric) distribution of RV planets, yet with low mutual inclinations imposed. Since we theoretically expect large eccentricities and large inclinations to go hand-in-hand, a tendency for planetary systems to be eccentric but flat would be surprising.

\acknowledgments We are grateful to the referee for a helpful report. We thank David Hogg, Gongjie Li, Katherine Deck, Joshua Carter, Guillaume H\'ebrard, Boas Katz, Yoram Lithwick, Smadar Naoz, Eugene Chiang, Scott Tremaine, Ellen Price, Leslie Rogers, Eric Ford, Cristobal Petrovich, and Doug Lin for helpful discussions. R.I.D. gratefully acknowledges the Miller Institute for Basic Research in Science, University of California Berkeley. J.A.J. is grateful for the generous grant support provided by the Alfred P. Sloan and David \& Lucile Packard foundations. D.F.M. is supported by NASA under grant NNX12AI50G and the National Science Foundation under grant IIS-1124794. DH acknowledges support by an appointment to the NASA Postdoctoral Program at Ames Research Center administered by Oak Ridge Associated Universities, and NASA Grant NNX14AB92G issued through the \kep Participating Scientist Program. This work benefited from the Summer Program on Modern Statistical and Computational Methods for Analysis of \kep Data, held at SAMSI, Research Triangle Park, NC in June 2013.

This paper includes data collected by the \kep mission. Funding for the \kep mission is provided by the NASA Science Mission directorate. We are grateful to the \kep Team for their extensive efforts in producing such high quality data. Some of the data presented in this paper were obtained from the Multimission Archive at the Space Telescope Science Institute (MAST). STScI is operated by the Association of Universities for Research in Astronomy, Inc., under NASA contract NAS5-26555. Support for MAST for non-HST data is provided by the NASA Office of Space Science via grant NNX09AF08G and by other grants and contracts.

We are very grateful to Geoff Marcy and Howard Isaacson for contributing to the radial velocity observations of Kepler-419. J.A.J. is grateful for Keck/HIRES time allocated through the Caltech Time Allocation Committee for some of the spectra used herein. The spectroscopic and radial-velocity measurements presented herein were obtained at the W.M. Keck Observatory, which is operated as a scientific partnership among the California Institute of Technology, the University of California and the National Aeronautics and Space Administration. The Observatory was made possible by the generous financial support of the W.M. Keck Foundation. We gratefully acknowledge the efforts and dedication of the Keck Observatory staff, especially Scott Dahm, Greg Doppman, Hien Tran, and Grant Hill for support of HIRES and Greg Wirth for support of remote observing.  We extend special thanks to those of Hawai`ian ancestry on whose sacred mountain of Mauna Kea we are privileged to be guests.  Without their generous hospitality, the Keck observations presented herein would not have been possible.

This paper uses observations obtained with facilities of the Las Cumbres Observatory Global Telescope. The Byrne Observatory at Sedgwick (BOS) is operated by the Las Cumbres Observatory Global Telescope Network and is located at the Sedgwick Reserve, a part of the University of California Natural Reserve System.

\appendix

\section{Two alternative spectroscopic analyses}
\label{app:spec}

{ In addition to the spectroscopic analysis described in Section \ref{sec:sme}, we extract the stellar properties from the spectrum using two approaches described below.  All three approaches are in agreement.}

Our second approach is to use SME Version 288 with the original \citet{2005V} spectral intervals, line list, and free parameters. This yields the stellar parameters in Table 1, Section \ref{sec:star}, which are consistent with results from our first approach, but with larger uncertainties. To estimate uncertainties, we perturb and temporarily fix one free parameter at a time, solving for the remaining free parameters. We then calculate the standard deviation of each derived parameter for all fits with reduced chi-squared less than the minimum value (3.39) plus 1. This approach \citep{1976A} provides the crude, but practical, uncertainty estimates in Table 1. The uncertainties are a few times larger than reported in \citet{2005V}, mainly because their line list provides a weaker gravity constraint for stars as warm as Kepler-419.
 
Our final approach is to use the stellar parameter classification (SPC) tool developed by \citet{2012BL}. In the SPC approach, the observed spectrum is cross-correlated against a collection of synthetic spectrum generated from a collection of sets of stellar parameters ($\teff,\log g,$ metal abundance relative to solar ([m/H]), and $v\sin i$). Based on the comparisons of SPC to other approaches performed by \citet{2012T}, we add 59 K and 0.062 dex in quadrature to the formal uncertainties in $\teff$ and [M/H] respectively.

\section{Fits to transit lights}
\label{app:fits}
{ For the fits using {\rm TAP}}, we use the \kep pre-search data conditioned (PDC) flux \citep{2010J,2010JC,2010T,2012S,2012SS}, which has been detrended for instrumental effects by removing systematic variations present in stars located nearby on the detector using cotrending basis vectors. To further detrend, we divide the PDC flux into chunks split by the observing quarter and/or a jump in the flux at the level of 5\% or higher. We smooth each chunk using a running median filter of width of 15 hours.We experiment with the width of the filter to ensure that it does not distort the transit depth. We discard the first and last 7.5 hours of each chunk. For each transit, we trim the total light curve (in and out of transit data) to 128 data points (2.7 days) for long cadence data and 4096 data points (2.8 days) for short cadence data). We thereby ensure that the number of data points is $2^N$ so that we make use of the \citet{2009C} wavelet likelihood (CW09 hereafter) without excessive zero-padding, as we have found that zero-padding can sometimes artificially decrease the uncertainties.

\begin{deluxetable}{rrlrl}
\tabletypesize{\footnotesize}%
\tablecaption{Mid epoch of transit [BJD-2454833] \label{tab:lcttv}}
\tablewidth{0pt}
\tablehead{
\colhead{Parameter}    & \colhead{Value\tablenotemark{a}}}
\startdata
												& {\tt TAP}\tablenotemark{b}		&				& GP\tablenotemark{c,d} \\
\hline
\\
$T_1$ [days]     				&126.3308&$^{+0.0010}_{-0.0009}$ 	&126.3305&$\pm0.0008$ 	\\\\
$T_2$ [days]  							&196.0606&$\pm$0.0006  		&196.0605&$\pm$0.0006 				\\\\
$T_3$ [days]  							&265.7661&$\pm$0.0006  		&265.7661&$\pm$0.0006  \\\\
$T_4$ [days]   							&335.5766&$\pm$0.0006  		&335.5762&$\pm$0.0006 \\\\
$T_5$ [days]  							&405.3154&$\pm$0.0006  		&405.3151&$\pm$0.0006  \\\\
 $T_6$ [days]  							&475.0077&$^{+0.0015}_{-0.0030}$ 		&475.0076&$^{+0.0019}_{-0.0045}$ 				\\\\
$T_7$ [days]  							&544.7262&$\pm$0.0006  		&544.7261&$\pm$0.0005 \\\\
$T_8$ [days] 							&614.4561&$\pm$0.0006  		&614.4558&$^{+0.0006}_{-0.0005}$\\\\ 
$T_9$ [days] 							&684.1878&$^{+0.0005}_{-0.0006}$  &684.1878&$^{+0.0006}_{-0.0005}$  \\\\
$T_{10}$ [days] 							&753.9190&$^{+0.0007}_{-0.0008}$  	&753.9199&$^{+0.0014}_{-0.0012}$  \\\\
$T_{11}$ [days] 							&823.6434&$\pm$0.0004  					&823.6435&$\pm$0.0003 \\\\ 
$T_{12}$ [days] 							&893.3498&$\pm$0.0004  					&893.3503&$\pm$0.0003 \\\\
$T_{13}$ [days] 							&963.0391&$\pm$0.0005  					&963.0387&$\pm$0.0004\\\\
$T_{14}$ [days] 							&1102.6198&$\pm$0.0004  					&1102.6201&$\pm$0.0003\\\\
$T_{15}$ [days] 							&1172.3055&$\pm$0.0004  					&1172.3056&$\pm$0.0003  \\\\
$T_{16}$ [days] 							&1242.0125&$\pm$0.0004  					&1242.0121&$^{+0.0004}_{-0.0003}$  \\\\
$T_{17}$ [days] 							&1311.7272&$\pm$0.0005  					&1311.7271&$\pm$0.0003  \\\\
$T_{18}$ [days] 							&1381.4428&$^{+0.0004}_{-0.0005}$  		&1381.4428&$\pm$0.0004  \\\\
$T_{19}$ [days] 							&1451.1566&$\pm$0.0005  				&1451.1567&$\pm$0.0003 	\\\\
$T_{20}$ [days] 							&1520.8606&$\pm$0.0004  				&1520.8602&$\pm$0.0003 	\\\\
$T_{21}$ [days] 							&1590.5432&$\pm$0.0006  				&1590.5431&$\pm0.0005$  \\\\
\enddata
\tablenotetext{a}{The uncertainties represent the 68.3\% confidence interval of the posterior distribution.}
\tablenotetext{b}{{\tt TAP} software by \citet{2011G}. Uses CW09 wavelet likelihood.}
\tablenotetext{c}{Daniel Foreman-Mackey et al., in prep}
\tablenotetext{d}{\citet{2013FH}}
\end{deluxetable}

\begin{deluxetable}{rrrrrrr}
\tabletypesize{\footnotesize}%
\tablecaption{Transit impact parameters \label{tab:lcb}}
\tablewidth{0pt}
\tablehead{
\colhead{Parameter}    & \colhead{Value\tablenotemark{a}}}
\startdata
												&{\tt TAP}\tablenotemark{b}		&&				& GP\tablenotemark{c,d} \\
												&b&+unc$_b$&-unc$_b$&b&+unc$_b$&-unc$_b$\\\\
\hline
\\
					$b_{1}$ 				&0.09 &0.08&0.06	&0.06 &0.08&0.04\\
					$b_{2}$  				&0.08 &0.06&0.05	&0.07 &0.07&0.05\\
					$b_{3}$  				&0.10 &0.06&0.06	&0.09 &0.06&0.06\\									
					$b_{4}$  				&0.14 &0.05&0.07	&0.13 &0.06&0.08\\
					$b_{5}$  				&0.05 &0.05&0.03	&0.07 &0.05&0.05\\
					$b_{6}$  				&0.18 &0.16&0.12	&0.24 &0.18&0.15\\
					$b_{7}$  				&0.14 &0.05&0.07	&0.14 &0.05&0.07\\
					$b_{8}$  				&0.22 &0.04&0.05	&0.20 &0.04&0.05\\
					$b_{9}$  				&0.12 &0.06&0.07	&0.11 &0.05&0.07\\
					$b_{10}$  				&0.09 &0.07&0.06	&0.15 &0.09&0.09\\
					$b_{11}$  				&0.07 &0.05&0.05	&0.05 &0.04&0.04\\
					$b_{12}$  				&0.14 &0.05&0.07	&0.10 &0.05&0.05\\
					$b_{13}$  				&0.17 &0.05&0.06	&0.20 &0.03&0.04\\
					$b_{14}$  				&0.19 &0.03&0.04	&0.19 &0.03&0.04\\
					$b_{15}$  				&0.13 &0.05&0.06	&0.16 &0.03&0.06\\
					$b_{16}$  				&0.07 &0.06&0.05	&0.07 &0.05&0.05\\
					$b_{17}$  				&0.13 &0.05&0.07	&0.09 &0.05&0.05\\
					$b_{18}$  				&0.12 &0.05&0.07	&0.11 &0.04&0.06\\
					$b_{19}$  				&0.18 &0.04&0.05	&0.20 &0.03&0.03\\
					$b_{20}$  				&0.13 &0.05&0.06	&0.17 &0.03&0.04\\
					$b_{21}$  				&0.13 &0.06&0.07	&0.16 &0.04&0.05\\
\enddata
\tablenotetext{a}{The uncertainties represent the 68.3\% confidence interval of the posterior distribution.}
\tablenotetext{b}{{\tt TAP} software by \citet{2011G}. Uses CW09 wavelet likelihood.}
\tablenotetext{c}{Daniel Foreman-Mackey et al., in prep}
\tablenotetext{d}{\citet{2013FH}}
\end{deluxetable}
\clearpage
\section{Dynamical fits to subsets of the available data}
\label{app:ttvfits}
\subsection{Co-planar fit to the TTVs}
\label{app:coplanar}

In this subsection we fix the following parameters: $m_{b}$, $e_{b}$, $\omega_{b}$,  $i_{b}$, $\Omega_{b}$, $i_c$, and node $\Omega_c$. We fit for $P_{b}$, $M_{b}$, $m_c$,  $P_c$, $e_c$, $\omega_c$, and $M_c$. We take the transiting planet's mass from RV measurements (derived in Section \ref{sec:threed}; however, note that the TTVs are insensitive to the transiting planet's mass) and its eccentricity and periapse from RV measurements (which are consistent with the value derived from the photometry in Section \ref{sec:lc}). In Section \ref{sec:threed}, we simultaneously fit the RVs and TTVs, leading to similar results. We list the medians and 68.3\% confidence intervals of the derived posteriors in Table \ref{tab:fits}. 

\begin{deluxetable}{rrrrrrr}
\tabletypesize{\small}%
\tablecaption{Additional fits, for comparison to Table 
\ref{tab:bestfit}, for planet Parameters for Kepler-419b and Kepler-419-c at epoch BJD 2455809.4009671761741629. All orbital elements are Jacobian.\label{tab:fits}}
\tablewidth{0pt}
\tablehead{
\colhead{Parameter}    				& \colhead{Coplanar/TTVs } 		&\colhead{TTVs} 				&\colhead{+b} 					&\colhead{+RVs} 	&\colhead{ SPC\tablenotemark{a}	}}
\startdata
$m_\star (m_\odot)\tablenotemark{b}	$ 	&1.39$^{+0.08}_{-0.07}$			& 1.39$^{+0.08}_{-0.07}$			& 1.39$^{+0.08}_{-0.07}$			& 1.39 $^{+0.08}_{-0.07}$					& 1.36 $^{+0.08}_{-0.08}$ \\\\
$R_\star \tablenotemark{b}$			&							&							&1.59$^{+0.19}_{-0.19}$			&1.59$^{+0.19}_{-0.19}$					&$1.77^{+0.07}_{-0.09}$\\\\
$m_{b} (M_{\rm Jup})$				&2.6 (fixed)					&2.6 (fixed)					&2.6 (fixed)					& 2.5$\pm0.3$							& 2.5 $\pm$ 0.3\\\\
$P_{b}$ (days)						&69.7550$\pm$0.0003			&$69.7551\pm0.0005$		&$69.7548^{+0.0005}_{-0.0007}$		&$69.7547^{+0.0008}_{-0.0013}$				&$69.7545^{+0.0009}_{-0.0014}$\\\\
$a_{b}$ (AU) \tablenotemark{c}			&0.370$^{+0.007}_{-0.006}$		&0.370$^{+0.007}_{-0.006}$	&0.370$^{+0.007}_{-0.006}$			&0.370$^{+0.007}_{-0.006}$					&0.368$^{+0.007}_{-0.006}$	\\\\
$e_{b}$							&0.823 (fixed)					&0.823 (fixed)					&0.823 (fixed)					&0.84$^{+0.03}_{-0.02}$						&0.839$^{+0.012}_{-0.014}$\\\\
$\omega_{b}(^\circ)$					&95.495 (fixed)					&95.495 (fixed)				&95.495 (fixed)						&94.2$^{0.9}_{-1.0}$							&96.5$^{+2.0}_{-0.7}$\\\\
$M_{b}(^\circ)$					&$68.6489^{+0.0018}_{-0.0018}$		&$68.650^{+0.002}_{-0.002}$	&$68.650^{+0.002}_{-0.002}$			&$68.75^{+0.04}_{-0.05}$						&$68.63^{+0.05}_{-0.08}$\\\\
$i_{b} (^\circ)$					&90 (fixed)						&90 (fixed)					&$89.10^{+0.13}_{-0.14}$				&$89.0^{+0.2}_{-0.2}$						&88.9$^{+0.2}_{-0.2}$\\\\		
$\Omega_{b}(^\circ)$				&0 (fixed)						&0 (fixed)						&0 (fixed)						& 0 (fixed)					
	& 0 (fixed)\\\\
$m_{c} (M_{\rm Jup})$				&$7.3\pm0.4$					&$7.4\pm0.4$					&$7.4\pm0.4$					&$7.3\pm0.4$							&$7.2^{+0.4}_{-0.4}$\\\\
$P_{c} (days)$						&$675.55\pm0.09$				&$675.52\pm0.10$				&$675.51\pm0.10$				&$675.45^{+0.13}_{-0.14}$				&$675.44^{+0.12}_{-0.12}$\\\\
$a_{c}$ (AU) 		\tablenotemark{c}				&1.68$\pm$0.03	&1.68$\pm$0.03	&1.68$\pm$0.03				&1.68$\pm$0.03							&1.67$\pm$0.03\\\\
$e_{c}$							&$0.1846^{+0.0008}_{-0.0007}$	&$0.1851^{+0.0008}_{-0.0008}$	&$0.1851^{+0.0008}_{-0.0008}$	&$0.184^{+0.002}_{-0.003}$				&$0.183^{+0.002}_{-0.002}	$\\\\
$\omega_{c} (^\circ)$				&$275.9^{+0.6}_{-0.6}$			&$275.7^{+0.5}_{-0.5}$			&$275.8^{+0.5}_{-0.6}$			&$274.7^{+0.9}_{-1.1}$					&$276.6^{+1.5}_{-1.0}$\\\\
$M_{c} (^\circ)$						&$345.0^{+0.4}_{-0.3}$			&$345.0\pm0.3$				&$345.0^{+0.3}_{-0.3}$			&$344.9^{+0.4}_{-0.3}$					&$345.2\pm0.4$\\\\
$\Omega_{c} (^\circ)$				&0 (fixed)						&$1^{+9}_{-9}$					&$-3^{+12}_{-9}$				&$3^{+12}_{-14}$						&$3^{+18}_{-13}$\\\\
$i_{c} (^\circ)$\						&90 (fixed)					&$90^{+2}_{-2}$			&$90^{+2}_{-3}$			&$89^{+3}_{-3}$						&$88^{+3}_{-2}$\\\\		
$i_{\rm mut} (^\circ)$					&0 (fixed)						&$7^{+5}_{-4}$					&$9^{+4}_{-5}$					&$10^{+7}_{-6}$							&$10^{+12}_{-6}$\\\\
$99\%$ $i_{\rm mut} (^\circ)$			&							&20							&20							&22										&29\\\\
$\omega_{b}- \omega_{c} (^\circ)$		&179.7$^{+0.6}_{-0.6}$			&179.9$^{+0.5}_{-0.5}$			&179.8$^{+0.6}_{-0.5}$				&179.6$^{+0.6}_{-0.6}$					&180.2$^{+0.8}_{-0.7}$\\\\
$\varpi_{b}- \varpi_{c} (^\circ)$		&179.7$^{+0.6}_{-0.6}$				&180$^{+9}_{-9}$				&182$^{+9}_{-12}$				&176$^{+14}_{-12}$							&177$^{+13}_{-17}$\\\\
Systemic offset (m/s)					&							&							&							&$-34\pm10$								&$-28\pm 10$\\\\
\enddata
\tablenotetext{a} {Same as fit in Table \ref{tab:bestfit} but using stellar parameters from SPC, Section 2 of \ref{tab:star}.}
\tablenotetext{b} {Posterior from Section \ref{sec:star} imposed as a prior.}
\tablenotetext{c} {Derived from stellar mass and orbital period posteriors.}
\end{deluxetable}
\clearpage

\subsection{Non-coplanar fit to the TTVs}
\label{app:TTVsfit}
Next we relax the assumption of coplanarity. In Appendix \ref{app:ttvs}, we argue that the component of the perturber's orbit in the direction of the highly-eccentric transiting planet's apoapse dominates the TTV signal and that, based on the geometry of this particular system, the TTV signal depends only weakly on the perturber's node $\Omega_c$. The perturber's line-of-sight inclination $i_c$ is very well-constrained because, as we show in Section \ref{sec:threed}, the inner planet's major axis happens to lie nearly along the line of sight (which is not surprising because its transit probability is highest at periapse). Therefore the mutual inclination depends mostly on the longitude node $\Omega_c$, which is not as well constrained. (See Appendix \ref{app:spec} for further details.)

Because the dependence of the TTV signal on $\Omega_c$ is relatively week and is multi-modal, we search for the global minimum by finding the best-fitting solution for every value of $\Omega_c$ in increments of 1 degree. We find this best-fitting solution using the Levenberg-Mardquart algorithm, implemented in {\tt mpfit} \citep{alias2009Markwardt}. However, to account for the skewness of the TTV posteriors (and later impact parameter posteriors), we use a modified residual that mimics one drawn from an asymmetric normal distribution. Instead of supplying the algorithm with an array of $-(m_i-y_i)/\sigma_i$, for which  $y_i$ is the ith data point, $m_i$ is the model, and $\sigma_i$ is the uncertainty (the negative sign is because the algorithm uses a negative residual), we supply it with:
\begin{eqnarray}
\label{eqn:ourchi}
-\frac{m_i - y_i}{|m_i-y_i|} \sqrt{\left(\frac{m_i - y_i}{\sigma_i^-}\right)^2+ 2\ln\frac{\sigma_i^-}{\sigma_{\rm min}}} & , m_i < y_i \nonumber\\
-\frac{m_i - y_i}{|m_i-y_i|} \sqrt{\left(\frac{m_i - y_i}{\sigma_i^+}\right)^2+2\ln\frac{\sigma_i^+}{\sigma_{\rm min}}} & , m_i > y_i \nonumber\\
\end{eqnarray}
\noindent for which $\sigma_i^+$ and  $\sigma_i^-$ are the difference between the median and the upper and lower limit (respectively) of the 68.3\% confidence interval, and $\sigma_{\rm min} = {\rm Min}[\sigma_i^+,\sigma_i^-]$.  When using the smaller of the upper vs. lower error bar, the expression reduces to $-(m_i-y_i)/\sigma_i$. When using the larger, the expression accounts for the different normalizations of the two halves of an asymmetric normal distribution.

\begin{figure}
\includegraphics{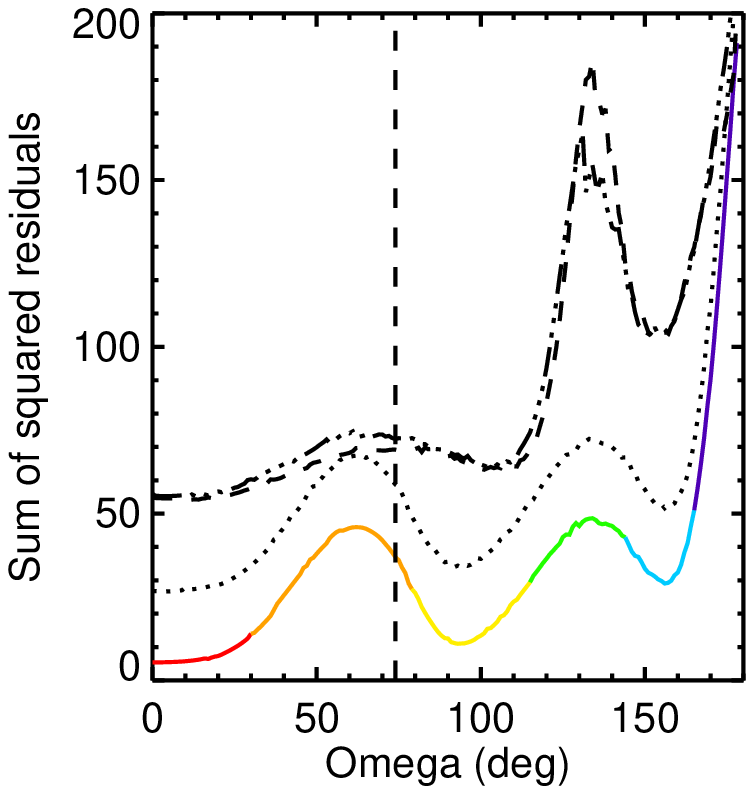}\\
\includegraphics{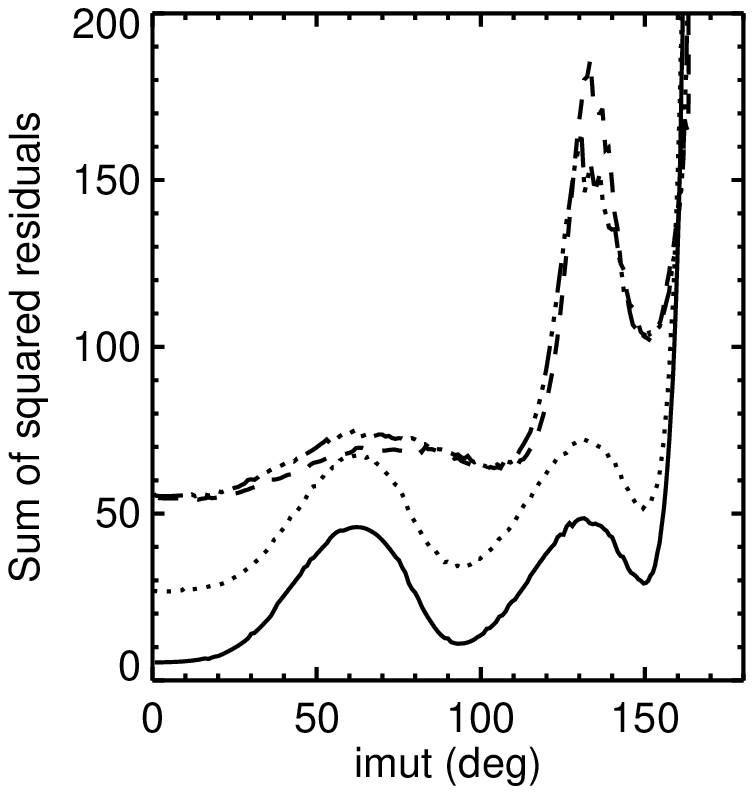}
\caption{\small Sum of squared residuals (summed square of Equation \ref{eqn:ourchi}) as a function of $\Omega_c$ (top) and total mutual inclination (bottom) for a fit to the TTVs only (solid line), TTVs and impact parameters $b$ (dotted line), TTVs/$b$/RVs (dashed line), and TTVs/$b$/RVs/$\rhocirc$ (dot-dashed line). The colors indicate ranges of $\Omega_c$ corresponding to those in Figure \ref{fig:omega}. The mutual inclination is very similar to $\Omega_c$ because $i_b$ is very tightly-constrained by the impact parameter and $i_c$ by the transit times (see Appendix \ref{app:ttvs}). The best-fitting solution is close to coplanar; there are two other local minima. The vertical dashed line indicates the $\Omega_c$ above which the inner planet collides with the star in long-term integrations (Appendix \ref{app:stable}). \label{fig:imut}}
\end{figure}

\begin{figure}
\includegraphics{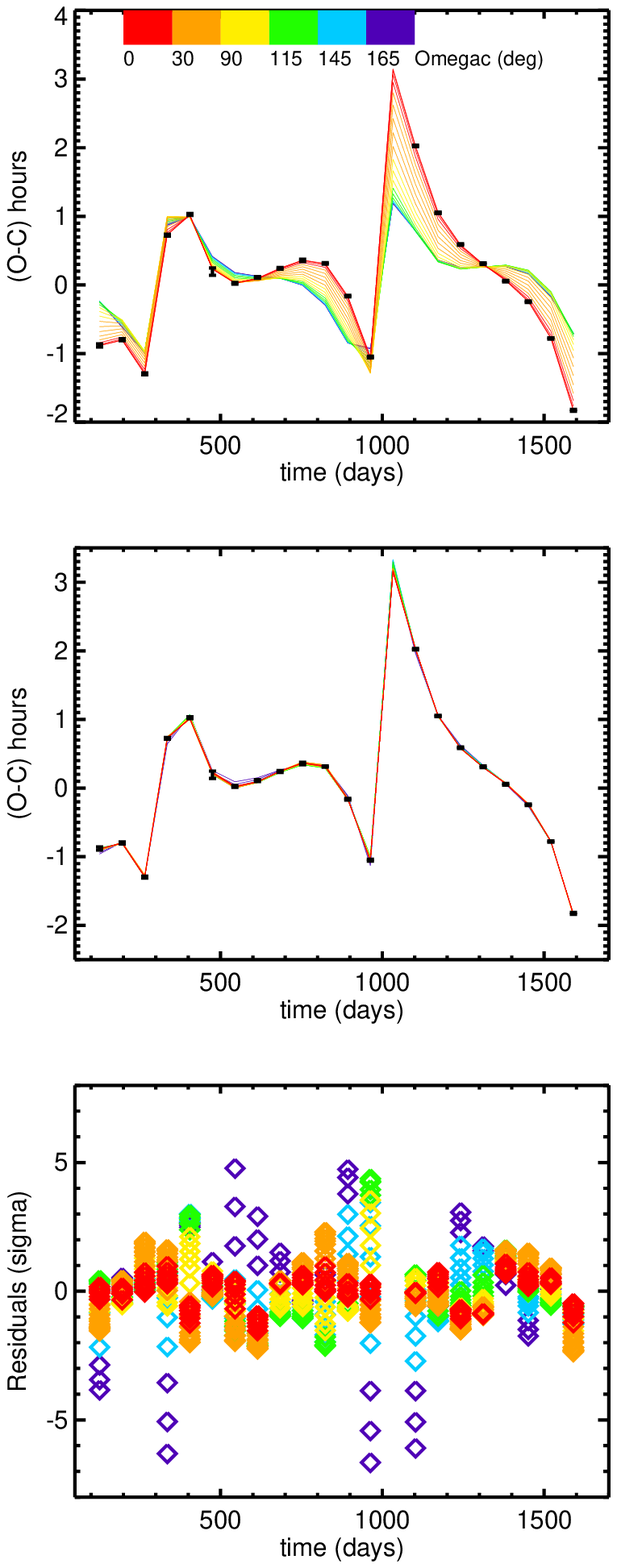}
\caption{\footnotesize Top: Best-fitting coplanar model (red) and, with other parameters fixed, varying the $\Omega_c$ from 0 (red) to $\pi$ (purple). Observations (with error bars) are over plotted in black. Middle: Same as above but with other parameters optimized for each $\Omega_c$. These models correspond to the solid line in Figure \ref{fig:imut}. Bottom: Residuals to the fits in the middle panel, in units of each data point's uncertainty. \label{fig:omega}}
\end{figure}

In the top panel of Figure \ref{fig:imut}, we plot this uncertainty-scaled sum of squared residuals (i.e. the summed square of Equation \ref{eqn:ourchi}) as a function of $\Omega_c$. Note that there are three local minima: one corresponding to a coplanar orbit, one to a polar orbit, and one to $\sim 150$ degree retrograde orbit. In the bottom panel, we plot the same quantity as a function of the mutual inclination. This bottom panel appears very similar to the top panel because $i_c$ is very tightly constrained by the transit times so the mutual inclination is mostly a function of $\Omega_c$ (see Appendix \ref{app:ttvs} for an explanation of why $i_c$ is better constrained than $\Omega_c$). { We include both the top and bottom panel to show that the uncertainty in the mutual inclination is almost entirely due to the uncertainty in $\Omega_c$.} Figure \ref{fig:omega} shows the subtle but detectable effect of $\Omega_c$ on the transit times. The top panel illustrates the change to the TTV signal caused by varying $\Omega_c$ only, the middle panel the change if we allow the other parameters to also vary to compensate, and the bottom panel the residuals to the middle panel. The color in the middle panel corresponds the colors of $\Omega_c$ in Figure \ref{fig:imut}.

Starting from the global minimum near $\Omega_c=0$ (Figure \ref{fig:imut}), we perform an MCMC fit, like that in Appendix \ref{app:coplanar} except allowing the Kepler-419's inclination and node to be free parameters. We report the medians of the posteriors and 68.3\% confidence intervals in column three of Table \ref{tab:fits}. From the TTVs alone, we constrain the mutual inclination to be { $7^{+5}_{-4}$ deg.}, consistent with coplanar and with a 99\% confidence upper limit of { $20^\circ$}.	

\subsection{Addition of the impact parameters}
\label{app:impactb}
In this section, we simultaneously model the measured mid transit epoch (Table \ref{tab:lcttv}) and the impact parameter $b$ of each transit (Table \ref{tab:lcb}). We repeat the fitting procedure from Appendix \ref{app:TTVsfit} except that we allow the inner planet's inclination to be a free parameter and the stellar radius to be a free parameter, with a prior on $R_\star$ imposed based on the posterior estimated in Section \ref{sec:star}. Although it is necessary to allow the impact parameter to be different for each transit to avoid a non-normally distributed $\rhocirc$ posterior (Figure \ref{fig:ttdur}, row 3), we caution that they may be caused by systematics not sufficiently accounted for in our red-noise models. In fact, no dynamical model in the posterior estimated from the fit to the TTVs in Appnedix \ref{app:TTVsfit} predicts a detectable variation in the impact parameter over the four years of observations. The impact parameters add no additional constraints to the planets' masses or orbital properties, except for the inner planet's inclination relative to the line of sight (i.e. which defines the impact parameter in combination with the stellar radius and the inner planet's eccentricity, periapse, and semi-major axis).  

\subsection{Supporting constraints from stability}
\label{app:stable}
\begin{figure}
\includegraphics{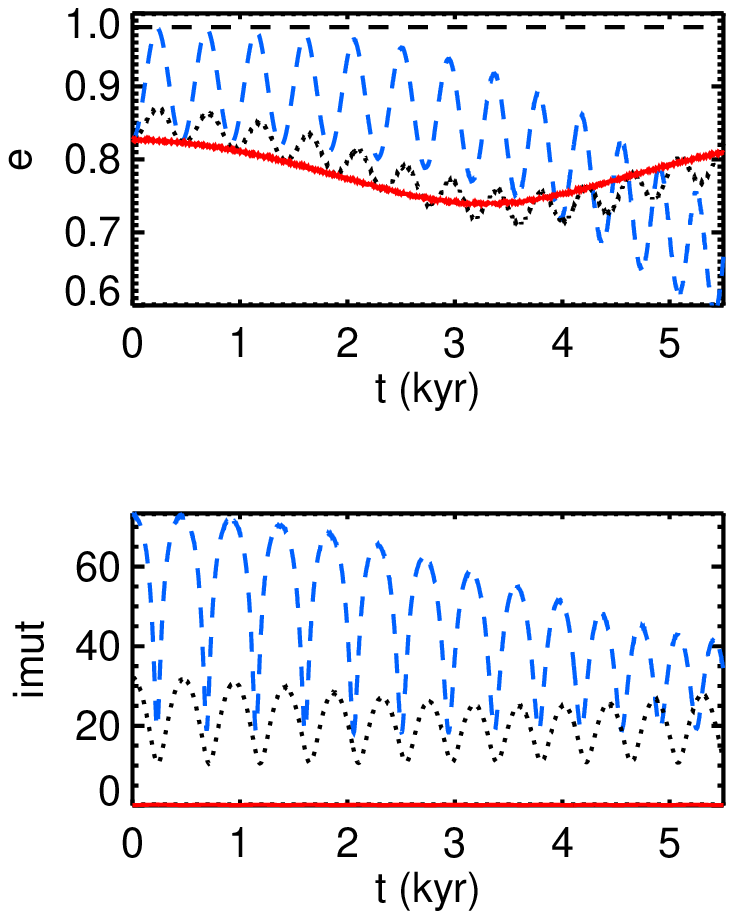}
\caption{Top: Evolution of the inner planet's eccentricity for $\Omega_c = 0$ (red), 32$^\circ$ (black), and 73$^\circ$ (cyan), taken from the fits to the TTVs/b (Figure \ref{fig:omega}). Above the black dashed line, the eccentricity is so high that the planet collides with the star. Bottom: Same for the evolution of the mutual inclination. \label{fig:int}}
\end{figure}

In addition to the better fit of the global minimum to the data, we favor the global minimum at low mutual inclination over the two local minima at large mutual inclinations (Fig. \ref{fig:omega}) because it is stable over a secular timescale. We integrate each of the 181 ``fixed node" solutions in Fig. \ref{fig:omega} for 0.5 Myr in {\tt Mercury6} \citep{1999C}, modified to include general-relativity and tidal precession following Section 2.1.1 of \citet{2010F}. In Figure \ref{fig:int}, we plot the evolution of the inner planet's eccentricity and mutual inclination for several characteristic cases. As we increase the mutual inclination, the amplitude of the secular eccentricity oscillation increases, particularly the shorter timescale mode that is coupled to the mutual inclination. For $74 < \Omega_c < 180$, the inner planet's eccentricity reaches such a high value that it collides with the star. Therefore we can rule out the two worse-fit, local minima solutions, giving further weight to the best-fitting, global minimum, low mutual inclination solution. 

\section{Spin-orbit alignment from projected rotational velocity}
\label{app:is}
\begin{figure}
\includegraphics{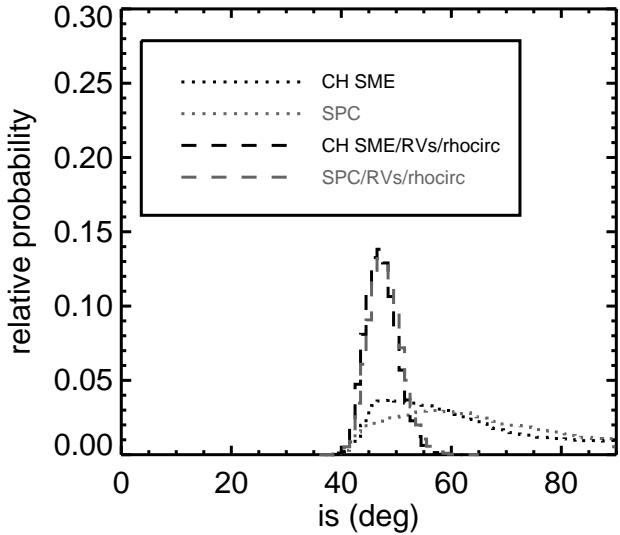}
\caption{Relative probability of the projected inclination of the host star's spin axis, for which $i_s=90^\circ$ is consistent with alignment with the planetary system, based on the stellar properties derived from Cargile, Hebb, et al. SME pipeline (black dotted line), combined with constraints from the photometry and RVs (black dashed line); same for SPC (gray dotted line, dshed line). \label{fig:is}}
\end{figure}

D12 measured a value of $i_s = 69^{+14}_{-17}$ degrees for the projected angle of the host star's spin axis, consistent with spin-orbit alignment $(i_s=i_{b})$ at the two-sigma level. Here we update that measurement, following the procedure of D12 Section 5.2 with an updated value of the stellar rotational period of $4.492 \pm 0.012$ days from \citet{2013M}, a projected rotational velocity $v \sin i_s$ from Section \ref{sec:star}, and stellar radius from Section \ref{sec:star} and Section \ref{sec:threed}. We plot the resulting posterior in Figure \ref{fig:is}, derived under four conditions: using the Cargile,~Hebb, et al. stellar radius (dotted black line), the SPC stellar radius (dotted gray line), and radius derived from fits to the TTVs, impact parameter, the light curve density, and RVs (Table \ref{tab:fits}, columns 4 and 5), making use of the Cargile,~Hebb, et al. stellar parameters (dashed black) and SPC stellar parameters (dotted black). While the first two cases are marginaly consistent with alignment, the third and fourth case are not. We measure the following four values for $i_s$ respectively (in degrees): $ 58^{+17}_{-10}, 62^{+16}_{-2}, 47\pm3,48\pm3$.

In other words, to be consistent with the measured $v \sin i$ and a stellar rotation axis perpendicular to the line of sight, the star's radius would need to be 1.3$R_\odot$ and its density would need to be 0.7 solar. Such a relatively large stellar density is marginally consistent with the stellar parameters derived from spectroscopy. However, it is inconsistent with the more precise stellar density measured from the light curve while accounting for the photo-eccentric effect based on constraints on $e$ and $\omega$ from the RVs.  

We consider whether the stellar density measured from the light curve could be biased. Blending, spots, TTVs, TDVs, and the planet's mass, as described by \cite{2014K}, can impact the measurement of the stellar density from the light curve at a precise level. We already allow for TTV and TDV in our models, and the effect of the planet's mass is too small to account for the discrepancy. Blending could cause the density to appear spuriously low but would require a star four times brighter than the one being transited \citep{2014K}. Given the constraints from adaptive optic imaging and radial-velocities (D12), such a blend is unlikely. 

Another possibility is that the uncertainties in the eccentricity we measured from the RVs are underestimated. To account for the discrepancy, the eccentricity would need to be about 0.7 instead of $0.83 \pm 0.01$ (Table \ref{tab:fits}), inconsistent with our uncertainties. However, to more { confidently} rule out this possibility, in the future we will better model the RV stellar noise, for example by correlating with the \kep photometry where available (e.g. \citealt{2012A}), to be sure our estimated parameters from the RVs and their uncertainties are as accurate as possible. For now, we conclude that there is some evidence that the entire system is misaligned from the host star's spin axis.

\section{Causes and uniqueness of the TTV signal}
\label{app:ttvs}
\begin{figure}
\includegraphics[width=.75\textwidth]{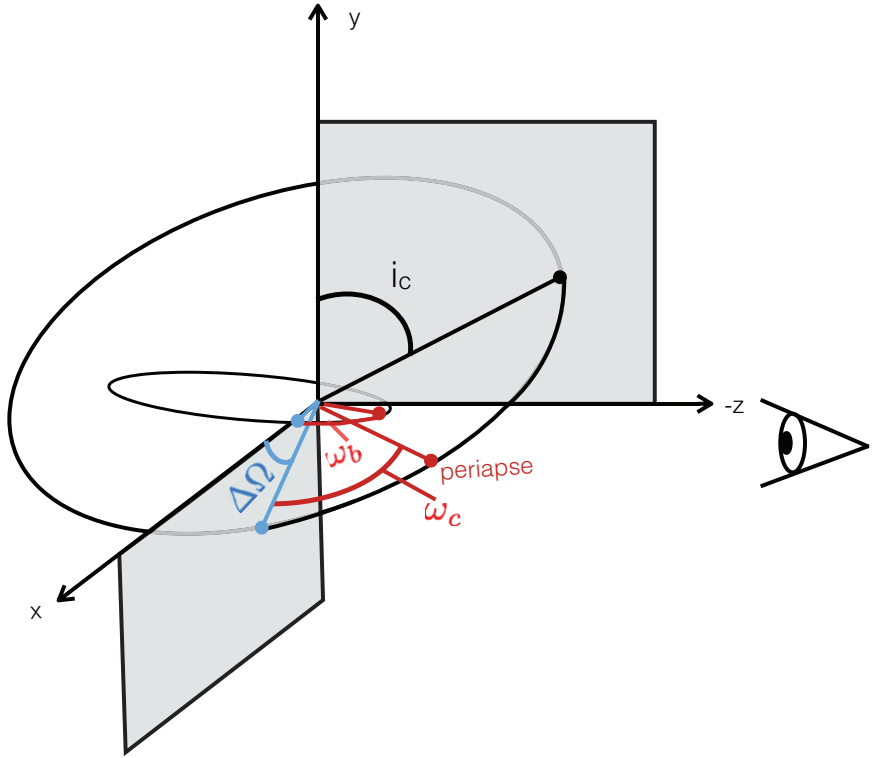}
\caption{Diagram of angles in sky frame. \label{fig:diagram}}
\end{figure}

\begin{figure}
\includegraphics{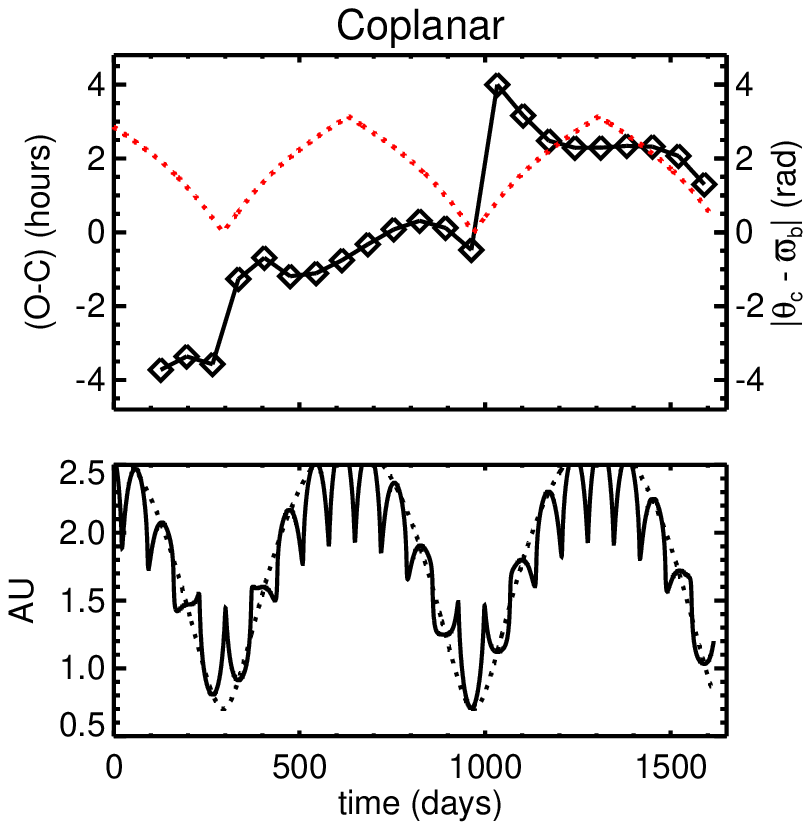} 
\includegraphics{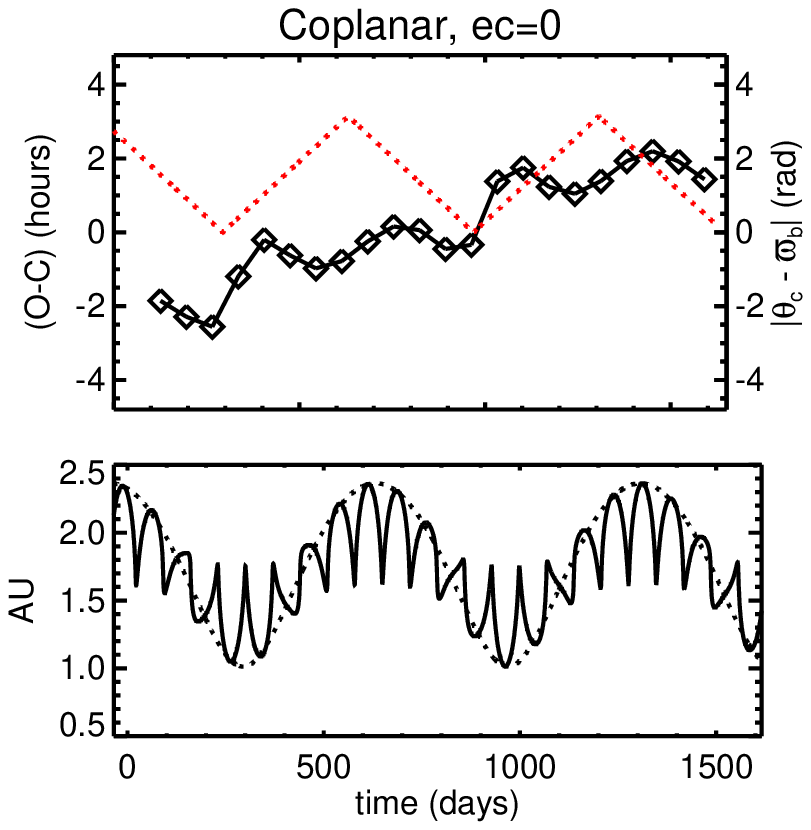}
\caption{Top: Kepler-419b's { deviation} from a linear ephemeris (solid black line, diamonds) and angular separation between Kepler-419c's position and Kepler-419b's apoapse (red dotted line; offset horizontally by half an orbital period of Kepler-419b toward decreasing time) for coplanar case with (left) and without (right) an eccentricity for Kepler-419c. Bottom: Separation of Kepler-419c from Kepler-419b's instantaneous position (solid) and from Kepler-419b's apoapse (dotted). Time axis is offset by half an orbital period of Kepler-419b (toward decreasing time) from above. (The reason for this plotted offset is that the deviation in the transit time is an integrated effect over an orbital period. Therefore, with this offset, the TTV signal is being affected by orbit changes plotted half a period on either side.).\label{fig:ttvecc}}
\end{figure}

Here we discuss in detail the cause of the transit timing variations and why it is possible to uniquely derive the properties of the perturbing companion without degeneracies. As discussed in Section \ref{sec:threed}, the TTVs have three potential contributors: the eccentricity of the inner planet (Kepler-419b), the eccentricity of the outer planet (Kepler-419c), and the mutual inclination. In Figure \ref{fig:diagram}, we show a diagram of angles. In Figure \ref{fig:ttvecc}, we plot the TTVs for a coplanar case with and without an eccentric orbit for Kepler-419c, selecting a constant linear ephemeris that highlights the kick that occurs when the long period Kepler-419c passes Kepler-419b's apoapse. During the kick, Kepler-419c reduces the effective central gravitational force felt by Kepler-419b, causing it to slow down and arrive late (positive O-C). Even with a coplanar, circular perturber, the TTVs are significant, but the kick is weaker because circular Kepler-419c is no longer at periapse when it passes Kepler-419b's apoapse.

\begin{figure}
\includegraphics{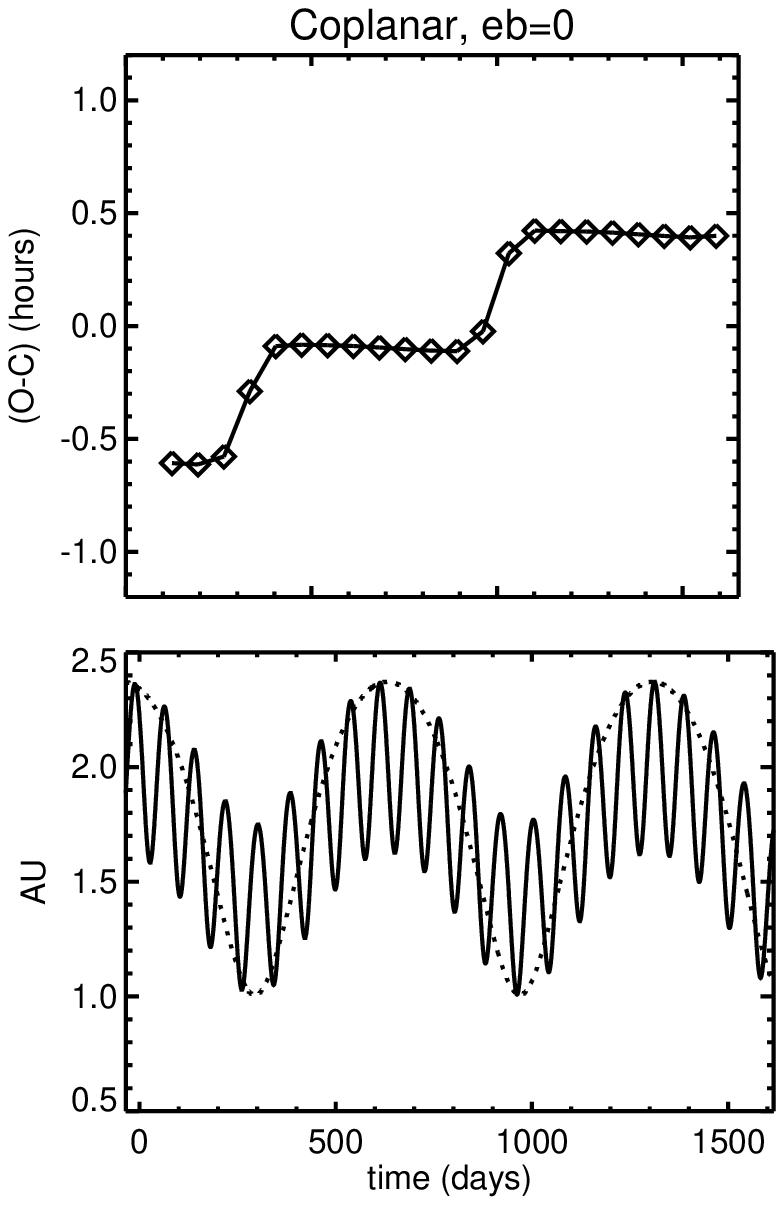} 
\includegraphics{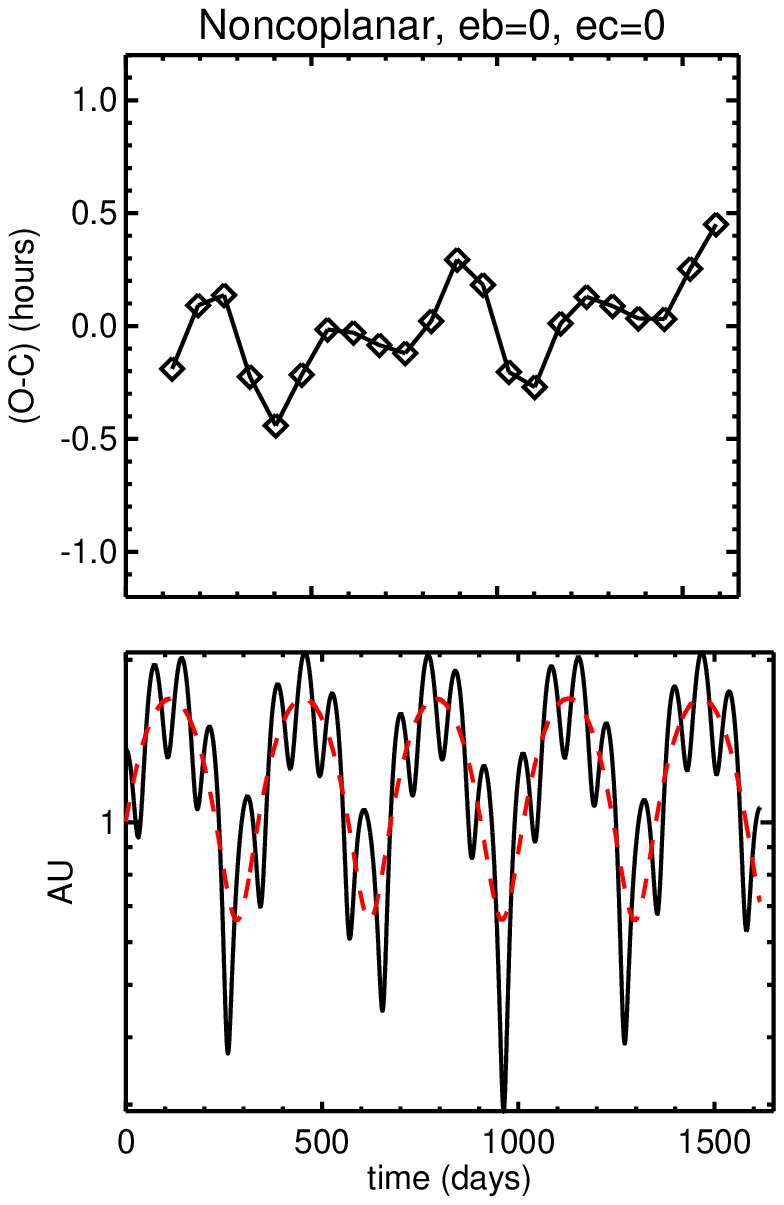} 
\caption{Top: Deviation from a linear ephemeris for coplanar case with $e_{b}=0$ (left) and non-coplanar case (right), for comparison to Figure \ref{fig:ttvecc} (note the different scale of the y-axis between Figure 11 and 12). Bottom: Separation of Kepler-419c from Kepler-419b's instantaneous position (solid, left) and from Kepler-419b when Kepler-419b is at conjunction with Kepler-419c's peripase (dotted, left); projected separation of Kepler-419c from Kepler-419b onto Kepler-419b's orbital plane (black solid, right) and position of Kepler-419c projected onto Kepler-419b's orbital plane (red dashed, right). Time axis is offset by half an orbital period of Kepler-419b from above.\label{fig:ttvecc2}}
\end{figure}

In Figure \ref{fig:ttvecc2} (left), we illustrate the effect of Kepler-419c's eccentricity; for illustrative purposes, we set Kepler-419b's eccentricity to zero. Without an eccentric orbit for Kepler-419b, the kicks are smaller (note the different scale of the y-axis { between Figure 11 and 12}) and the deviations from a linear ephemeris occur only near Kepler-419c's periapse passage. In the right panels of Figure \ref{fig:ttvecc2}, we illustrate the effect of mutual inclination; we set both planets' eccentricities to zero. The kicks now occur at twice Kepler-419c's orbital frequency, when it intersects Kepler-419b's orbital plane. The strongest kick is when the intersection occurs at conjunction. The different shape of the TTVs caused by Kepler-419c's eccentricity ({ Figure 12,} left) and mutual inclination ({ Figure 12,} right) allow us to constrain each of these two quantities without a strong degeneracy. Kepler-419b's eccentricity (Figure \ref{fig:ttvecc}, right) has some degeneracy with both effects, but we constrain Kepler-419b's eccentricity independently from the photoeccentric effect and RVs rather than measuring it from the TTVs.

The late arrival of Kepler-419b (Figures \ref{fig:ttvecc} and \ref{fig:ttvecc2}) is caused by a reduction in the effective central gravitational force felt by Kepler-419b as Kepler-419c pulls it away from the star. (This effect can also be conceptualized as a temporary reduction in the star's effective mass.) Therefore we can gain insight the constraints from the TTVs by examining the expression for the radial disturbing force per unit mass, $\bar{R}$, the radial component $({\bf \hat{r}})$ of the perturbing acceleration on the inner planet, $\delta{\bf\ddot{r}}$ (\citealt{2000M}, Equation 6.8). (Note that here we are not deriving an analytical expression for the TTVs, for which we would need to average over the orbit of the inner planet. Rather we are examining the force that causes the TTVs to gain insight into how the properties of the perturber affect the TTV signal. See \cite{2003BE} for analytical approximations to TTVs caused by an eccentric, inclined perturber.)

\begin{eqnarray}
\label{eqn:delta}
\bar{R}= \delta{\bf\ddot{r_b}} \cdot {\bf \hat{r_b}} = Gm'( \frac{{\bf r_c \cdot \hat{r_b}}-r_b}{|{\bf r_c-r_b}|^3}- \frac{\bf r_c \cdot \hat{r_b}}{r_c^3}) 
\end{eqnarray}
\noindent where $r'$ is the position of the outer planet, $G$ is the universal gravitational constant, and $m'$ is the mass of the perturbing planet.

The  term $\frac{1}{|{\bf r_c-r_b}|}$ can be expanded in Legrende polynomials (\citet{2000M}, 6.21) in powers of the separation ratio $(r_b/r_c)$:
\begin{eqnarray}
\frac{1}{|{\bf r_c-r_b}|} = \frac{1}{r_c} \sum_{l=0}^\infty \left(\frac{r_b}{r_c}\right)^l P_l({\bf \hat{r_b} \cdot \hat{r_c}}) =  \frac{1}{r_c} (1+ \sum_{l=1}^\infty \left(\frac{r_b}{r_c}\right)^l P_l({\bf \hat{r_b} \cdot \hat{r_c}}) )
\end{eqnarray}
\noindent so
\begin{eqnarray}
\label{eqn:expandcube}
\frac{1}{|{\bf r_c-r_b}|^3} = \frac{1}{r_c^3} \left( 1+\sum_{k=1}^\infty (^3_k) \left[\sum_{l=1}^\infty \left(\frac{r_b}{r_c}\right)^l P_l({\bf \hat{r_b} \cdot \hat{r_c}}) \right]^k\right)
\end{eqnarray}

Substituting Equation \ref{eqn:expandcube} into Equation \ref{eqn:delta}:
\begin{eqnarray}
\label{eqn:barr}
\bar{R}  = -\frac{Gm_c}{r_c^3} \left[-r_b + \left(r_c {\bf \hat{r_c} \cdot \hat{r_b}}-r_b\right) \left(\sum_{k=1}^\infty \left(^3_k\right) \left[\sum_{l=1}^\infty \left(\frac{r_b}{r_c}\right)^l P_l({\bf \hat{r_b} \cdot \hat{r_c}}) \right]^k\right)\right]
\end{eqnarray}

The first term of Equation \ref{eqn:barr} is independent of the planets' mutual inclination. Consequently, given our independent knowledge of the inner planet's separation $r_b$, we can measure the outer planet's mass $m_c$ (not just $m_c\sin i_c$ as in RV measurements) from the amplitude and eccentricity from time variation in $r_c$. If the outer planet's orbit were perfectly circular, we would still detect the outer planet's mass via the time variation of $r_b$ due to the inner planet's eccentricity. The second term allows us to measure the mutual inclination. Because each term in the sum has a different time dependence, the terms affect the signal in a non-degenerate way.

\begin{figure}
\includegraphics[width=5in]{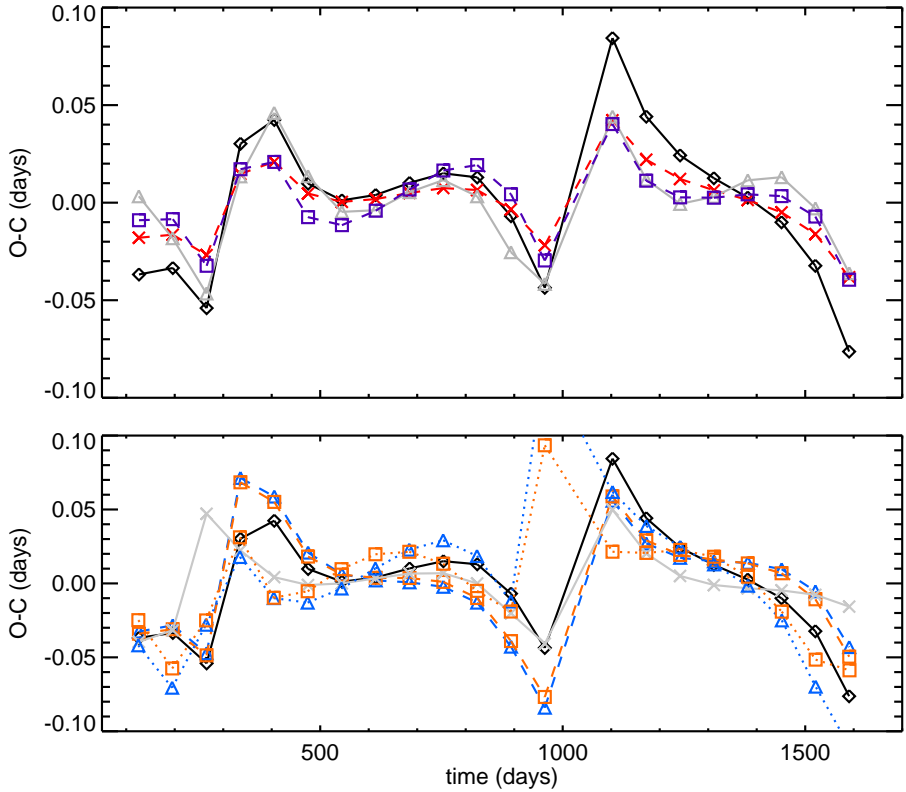}  
\caption{Top: Deviation from a linear ephemeris for: nominal coplanar case (black/solid/diamond) and modified by reducing the perturber's mass by a factor of 2 (gray/solid/triangle), decreasing 
$i_c$ from 90 to 67$^\circ$ (purple/dashed/square), and decrease $e_c$ from 0.18 to 0.07 (red/dotted/X). Although all three changes affect the amplitude, they affect the shape differently. Bottom: Same as above but modified by increasing the orbital period $P_c$ 
by 123 days (gray/solid/X), increasing the mean anomaly by 3 degrees (blue/triangle/dashed) and 46 degrees (blue/triangle/dotted), and increasing the argument of periapse by 5 degrees (orange/square/dashed) and 50 degrees (orange/square/dotted). Although all three changes affect the phase, they are distinguishable. See Figure \ref{fig:omega} for the subtle effect of $\Omega_c$ on the TTV signal. \label{fig:change}}
\end{figure}

Let us define the reference directions of the system so that the inner planet's node $\Omega_b=0$ and inclination $i_b=90$ (Figure \ref{fig:diagram}). Therefore the inner planet's position vector direction is ${\bf \hat{r_b}} = [u_{x,b},u_{y,b},u_{z,b}]$, where
\begin{eqnarray}
u_{x,b} =& \cos(f_b+\omega_b) \\
u_{y,b} =& 0 \\
u_{z,b} =& \sin(f_b+\omega_b) 
\end{eqnarray}
\noindent and $f_b$ is the mean anomaly and $\omega_b$ is the argument of periapse. The outer planet's position vector direction is ${\bf \hat{r_c}} = [u_{x,c},u_{y,c},u_{z,c}]$, where
\begin{eqnarray}
u_{x,c} =&  \cos\Omega_c \cos(f_c+\omega_c) -  \cos i_c \sin\Omega_c \sin(f_c+\omega_c)\\
u_{y,c} =&  \sin \Omega_cc \cos(f_c+\omega_c) +   \cos i_c \cos\Omega_c \sin(f_c+\omega_c)\\
u_{z,c}' =& \sin i_c\sin(f_c+\omega_c)
\end{eqnarray}
\noindent so $\hat{r_b} \cdot \hat{r_c} = u_{x,b}u_{x,c}+u_{z,b}u_{z,c}$. The magnitude of $u_{z,b}u_{z,c}$ strongly constraints $\sin i_c$. Furthermore, if the inner planet spends most of its time near apoapse and the perturbation is strongest there, the perturbation depends mostly on the component of the outer companion's position in the direction of the inner planet's apoapse. If the inner planet is transiting near periapse, as is the case for Kepler-419b, then the planet's major axis lies approximately along the z direction, and the perturbation depends primarily on the $u_{z,c}$ component of the outer planet's position. Since the $u_{z,c}$ component is only a function of the outer planet's inclination $i_c$, not its node $\Omega_c$, when a planet transits near periapse one gets a tight constraint on its companion $i_c$ but a weaker constraint on $\Omega_c$. This is the case for Kepler-419b. Figure \ref{fig:omega} illustrates the relatively weak dependence of the TTV signal on $\Omega_c$. Figure \ref{fig:change} shows the much stronger effect of the other parameters on the TTV signal and their lack of degeneracy.

\bibliography{./pehj4} \bibliographystyle{apj}

\begin{thebibliography}{49}
\expandafter\ifx\csname natexlab\endcsname\relax\def\natexlab#1{#1}\fi

\bibitem[{{Agol} {et~al.}(2005){Agol}, {Steffen}, {Sari}, \&
  {Clarkson}}]{2005A}
{Agol}, E., {Steffen}, J., {Sari}, R., \& {Clarkson}, W. 2005, \mnras, 359, 567

\bibitem[Aigrain et al.(2012)]{2012A} Aigrain, S., Pont, F., 
\& Zucker, S.\ 2012, \mnras, 419, 3147 

\bibitem[{{Albrecht} {et~al.}(2012){Albrecht}, {Winn}, {Johnson}, {Howard},
  {Marcy}, {Butler}, {Arriagada}, {Crane}, {Shectman}, {Thompson}, {Hirano},
  {Bakos}, \& {Hartman}}]{2012AW}
{Albrecht}, S., {Winn}, J.~N., {Johnson}, J.~A., et al. 2012, \apj,
  757, 18
  
  \bibitem[Ambikasaran et al.(2014)]{2014A} Ambikasaran, S., 
Foreman-Mackey, D., Greengard, L., Hogg, D.~W., 
\& O'Neil, M.\ 2014, arXiv:1403.6015 

\bibitem[Avni(1976)]{1976A} Avni, Y.\ 1976, \apj, 210, 642 

\bibitem[{{Ballard} {et~al.}(2011){Ballard}, {Fabrycky}, {Fressin},
  {Charbonneau}, {Desert}, {Torres}, {Marcy}, {Burke}, {Isaacson}, {Henze},
  {Steffen}, {Ciardi}, {Howell}, {Cochran}, {Endl}, {Bryson}, {Rowe}, {Holman},
  {Lissauer}, {Jenkins}, {Still}, {Ford}, {Christiansen}, {Middour}, {Haas},
  {Li}, {Hall}, {McCauliff}, {Batalha}, {Koch}, \& {Borucki}}]{2011B}
{Ballard}, S., {Fabrycky}, D., {Fressin}, F., et al. 2011, \apj, 743, 200


\bibitem[Bessell(1990)]{1990B} Bessell, M.~S.\ 1990, \pasp, 
102, 1181 

\bibitem[{{Borkovits} {et~al.}(2003){Borkovits}, {{\'E}rdi},
  {Forg{\'a}cs-Dajka}, \& {Kov{\'a}cs}}]{2003BE}
{Borkovits}, T., {{\'E}rdi}, B., {Forg{\'a}cs-Dajka}, E., \& {Kov{\'a}cs}, T.
  2003, \aap, 398, 1091

\bibitem[{{Brown} {et~al.}(2013){Brown}, {Baliber}, {Bianco}, {Bowman},
  {Burleson}, {Conway}, {Crellin}, {Depagne}, {De Vera}, {Dilday}, {Dragomir},
  {Dubberley}, {Eastman}, {Elphick}, {Falarski}, {Foale}, {Ford}, {Fulton},
  {Garza}, {Gomez}, {Graham}, {Greene}, {Haldeman}, {Hawkins}, {Haworth},
  {Haynes}, {Hidas}, {Hjelstrom}, {Howell}, {Hygelund}, {Lister}, {Lobdill},
  {Martinez}, {Mullins}, {Norbury}, {Parrent}, {Paulson}, {Petry}, {Pickles},
  {Posner}, {Rosing}, {Ross}, {Sand}, {Saunders}, {Shobbrook}, {Shporer},
  {Street}, {Thomas}, {Tsapras}, {Tufts}, {Valenti}, {Vander Horst}, {Walker},
  {White}, \& {Willis}}]{2013BBB}
{Brown}, T.~M., {Baliber}, N., {Bianco}, F.~B., et al. 2013, \pasp, 125, 1031

\bibitem[Buchhave et al.(2012)]{2012BL} Buchhave, L.~A., 
Latham, D.~W., Johansen, A., et al.\ 2012, \nat, 486, 375 

\bibitem[{{Butler} {et~al.}(1996){Butler}, {Marcy}, {Williams}, {McCarthy},
  {Dosanjh}, \& {Vogt}}]{1996B}
{Butler}, R.~P., {Marcy}, G.~W., {Williams}, E., et al. 1996, \pasp, 108, 500

\bibitem[Campante et al.(2014)]{2014CC} Campante, T.~L., 
Chaplin, W.~J., Lund, M.~N., et al.\ 2014, \apj, 783, 123 

\bibitem[{{Carter} \& {Winn}(2009)}]{2009C}
{Carter}, J.~A., \& {Winn}, J.~N. 2009, \apj, 704, 51

\bibitem[{{Chambers}(1999)}]{1999C}
{Chambers}, J.~E. 1999, \mnras, 304, 793

\bibitem[Chaplin et al.(2011)]{2011C} Chaplin, W.~J., 
Kjeldsen, H., Bedding, T.~R., et al.\ 2011, \apj, 732, 54 

\bibitem[Chatterjee et al.(2008)]{2008CF} Chatterjee, S., 
Ford, E.~B., Matsumura, S., \& Rasio, F.~A.\ 2008, \apj, 686, 580 

\bibitem[Chiang et al.(2001)]{2001C} Chiang, E.~I., 
Tabachnik, S., \& Tremaine, S.\ 2001, \aj, 122, 1607 

\bibitem[{{Chiang} \& {Murray}(2002)}]{2002C}
{Chiang}, E.~I., \& {Murray}, N. 2002, \apj, 576, 473

\bibitem[{{Cumming} {et~al.}(2008){Cumming}, {Butler}, {Marcy}, {Vogt},
  {Wright}, \& {Fischer}}]{2008C}
{Cumming}, A., {Butler}, R.~P., {Marcy}, G.~W., et al. 2008, \pasp, 120, 531

\bibitem[{{Dawson} \& {Johnson}(2012)}]{2012DJ}
{Dawson}, R.~I., \& {Johnson}, J.~A. 2012, \apj, 756, 122

\bibitem[{{Dawson} {et~al.}(2012{\natexlab{a}}){Dawson}, {Johnson}, {Morton},
  {Crepp}, {Fabrycky}, {Murray-Clay}, \& {Howard}}]{2012DJM}
{Dawson}, R.~I., {Johnson}, J.~A., {Morton}, T.~D., et al. 2012{\natexlab{a}} (D12), \apj,
  761, 163

\bibitem[Dawson 
\& Murray-Clay(2013)]{2013D} Dawson, R.~I., \& Murray-Clay, R.~A.\ 2013, \apjl, 767, L24 

\bibitem[{{Dawson} {et~al.}(2012{\natexlab{b}}){Dawson}, {Murray-Clay}, \&
  {Johnson}}]{2012DMJ}
{Dawson}, R.~I., {Murray-Clay}, R.~A., \& {Johnson}, J.~A. 2012{\natexlab{b}},
  ArXiv e-prints
  
\bibitem[Dong et al.(2014)]{2014D} Dong, S., Katz, B., 
\& Socrates, A.\ 2014, \apjl, 781, L5 

\bibitem[{{Dunhill} {et~al.}(2013){Dunhill}, {Alexander}, \&
  {Armitage}}]{2013DA}
{Dunhill}, A.~C., {Alexander}, R.~D., \& {Armitage}, P.~J. 2013, \mnras, 428,
  3072

\bibitem[{{Fabrycky}(2010)}]{2010F}
{Fabrycky}, D.~F. 2010, {Non-Keplerian Dynamics of Exoplanets}, ed. {Seager,
  S.}, 217--238

\bibitem[{{Foreman-Mackey} {et~al.}(2013){Foreman-Mackey}, {Hogg}, {Lang}, \&
  {Goodman}}]{2013FH}
{Foreman-Mackey}, D., {Hogg}, D.~W., {Lang}, D., \& {Goodman}, J. 2013, \pasp,
  125, 306
  
  \bibitem[Fulton et al.(2011)]{2011F} Fulton, B.~J., Shporer, 
A., Winn, J.~N., et al.\ 2011, \aj, 142, 84 

\bibitem[{{Gazak} {et~al.}(2012){Gazak}, {Johnson}, {Tonry}, {Dragomir},
  {Eastman}, {Mann}, \& {Agol}}]{2011G}
{Gazak}, J.~Z., {Johnson}, J.~A., {Tonry}, J., et al. 2012, Advances in Astronomy, 2012

\bibitem[Gibson et al.(2012)]{2012G} Gibson, N.~P., Aigrain, 
S., Roberts, S., et al.\ 2012, \mnras, 419, 2683 

\bibitem[{{Girardi} {et~al.}(2005){Girardi}, {Groenewegen}, {Hatziminaoglou},
  \& {da Costa}}]{2005G}
{Girardi}, L., {Groenewegen}, M.~A.~T., {Hatziminaoglou}, E., \& {da Costa}, L.
  2005, \aap, 436, 895
  
  \bibitem[{{Goldreich} \& {Sari}(2003)}]{2003G}
{Goldreich}, P., \& {Sari}, R. 2003, \apj, 585, 1024

\bibitem[{{Goldreich} \& {Tremaine}(1980)}]{1980G}
{Goldreich}, P., \& {Tremaine}, S. 1980, \apj, 241, 425

\bibitem[Guillochon et al.(2011)]{2011GR} Guillochon, J., 
Ramirez-Ruiz, E., \& Lin, D.\ 2011, \apj, 732, 74 

\bibitem[{{Howard} {et~al.}(2010){Howard}, {Johnson}, {Marcy}, {Fischer},
  {Wright}, {Bernat}, {Henry}, {Peek}, {Isaacson}, {Apps}, {Endl}, {Cochran},
  {Valenti}, {Anderson}, \& {Piskunov}}]{2010HJ}
{Howard}, A.~W., {Johnson}, J.~A., {Marcy}, G.~W., et al. 2010, \apj, 721, 1467

\bibitem[{{Howard} {et~al.}(2011){Howard}, {Johnson}, {Marcy}, {Fischer},
  {Wright}, {Henry}, {Isaacson}, {Valenti}, {Anderson}, \& {Piskunov}}]{2011H}
{Howard}, A.~W., {Johnson}, J.~A., {Marcy}, G.~W., et al. 2011, \apj, 726, 73

\bibitem[Huber et al.(2013)]{2013HC} Huber, D., Carter, J.~A., 
Barbieri, M., et al.\ 2013, Science, 342, 331 

\bibitem[Huber et al.(2013)]{2013H} Huber, D., Chaplin, 
W.~J., Christensen-Dalsgaard, J., et al.\ 2013, \apj, 767, 127 

\bibitem[Jenkins et al.(2010a)]{2010J} Jenkins, J.~M., 
Caldwell, D.~A., Chandrasekaran, H., et al.\ 2010, \apjl, 713, L87 

\bibitem[Jenkins et al.(2010b)]{2010JC} Jenkins, J.~M., 
Caldwell, D.~A., Chandrasekaran, H., et al.\ 2010, \apjl, 713, L120 

\bibitem[{{Johnson} {et~al.}(2012){Johnson}, {Gazak}, {Apps}, {Muirhead},
  {Crepp}, {Crossfield}, {Boyajian}, {von Braun}, {Rojas-Ayala}, {Howard},
  {Covey}, {Schlawin}, {Hamren}, {Morton}, {Marcy}, \& {Lloyd}}]{2012J}
{Johnson}, J.~A., {Gazak}, J.~Z., {Apps}, K.,et al. 2012, \aj, 143, 111

\bibitem[{{Johnson} {et~al.}(2010){Johnson}, {Howard}, {Marcy}, {Bowler},
  {Henry}, {Fischer}, {Apps}, {Isaacson}, \& {Wright}}]{2010JH}
{Johnson}, J.~A., {Howard}, A.~W., {Marcy}, G.~W., et al. 2010,
  \pasp, 122, 149

\bibitem[{{Juri{\'c}} \& {Tremaine}(2008)}]{2008J}
{Juri{\'c}}, M., \& {Tremaine}, S. 2008, \apj, 686, 603

\bibitem[{{Kaib} {et~al.}(2013){Kaib}, {Raymond}, \& {Duncan}}]{2013KR}
{Kaib}, N.~A., {Raymond}, S.~N., \& {Duncan}, M. 2013, \nat, 493, 381

\bibitem[{{Kenyon} \& {Bromley}(2008)}]{2008KB}
{Kenyon}, S.~J., \& {Bromley}, B.~C. 2008, \apjs, 179, 451

\bibitem[{{Kipping}(2013)}]{2013K}
{Kipping}, D.~M. 2013, \mnras, 435, 2152

\bibitem[Kipping(2014)]{2014K} Kipping, D.~M.\ 2014, \mnras, 
440, 2164 

\bibitem[Lega et al.(2013)]{2013L} Lega, E., Morbidelli, A., 
\& Nesvorn{\'y}, D.\ 2013, \mnras, 431, 3494 

\bibitem[Li et al.(2014)]{2014L} Li, G., Naoz, S., Kocsis, 
B., \& Loeb, A.\ 2014, \apj, 785, 116 

\bibitem[{{Lithwick} {et~al.}(2012){Lithwick}, {Xie}, \& {Wu}}]{2012LX}
{Lithwick}, Y., {Xie}, J., \& {Wu}, Y. 2012, \apj, 761, 122

\bibitem[{{McArthur} {et~al.}(2010){McArthur}, {Benedict}, {Barnes},
  {Martioli}, {Korzennik}, {Nelan}, \& {Butler}}]{2010M}
{McArthur}, B.~E., {Benedict}, G.~F., {Barnes}, R., et al. 2010, \apj, 715, 1203

\bibitem[McQuillan et al.(2013)]{2013M} McQuillan, A., Mazeh, 
T., \& Aigrain, S.\ 2013, \apjl, 775, L11 

\bibitem[Mandel 
\& Agol(2002)]{2002M} Mandel, K., \& Agol, E.\ 2002, \apjl, 580, L171 

\bibitem[{{Markwardt}(2009)}]{alias2009Markwardt}
{Markwardt}, C.~B. 2009, in Astronomical Society of the Pacific Conference
  Series, Vol. 411, Astronomical Data Analysis Software and Systems XVIII, ed.
  D.~A. {Bohlender}, D.~{Durand}, \& P.~{Dowler}, 251

\bibitem[Michtchenko 
\& Malhotra(2004)]{2004M} Michtchenko, T.~A., \& Malhotra, R.\ 2004, \icarus, 168, 237 

\bibitem[{{Mikkola}(1997)}]{1997M}
{Mikkola}, S. 1997, Celestial Mechanics and Dynamical Astronomy, 67, 145

\bibitem[{{Morton}(2012)}]{2012M}
{Morton}, T.~D. 2012, \apj, 761, 6

\bibitem[{{Morton} \& {Johnson}(2011)}]{2011M}
{Morton}, T.~D., \& {Johnson}, J.~A. 2011, \apj, 729, 138

\bibitem[{{Murray} \& {Dermott}(2000)}]{2000M}
{Murray}, C.~D., \& {Dermott}, S.~F. 2000, {Solar System Dynamics}

\bibitem[{{Naoz} {et~al.}(2011){Naoz}, {Farr}, {Lithwick}, {Rasio}, \&
  {Teyssandier}}]{2011NF}
{Naoz}, S., {Farr}, W.~M., {Lithwick}, Y., et al.
  2011, \nat, 473, 187

\bibitem[{{Naoz} {et~al.}(2012){Naoz}, {Farr}, \& {Rasio}}]{2012N}
{Naoz}, S., {Farr}, W.~M., \& {Rasio}, F.~A. 2012, \apjl, 754, L36

\bibitem[{{Nesvorn{\'y}} {et~al.}(2012){Nesvorn{\'y}}, {Kipping}, {Buchhave},
  {Bakos}, {Hartman}, \& {Schmitt}}]{2012NK}
{Nesvorn{\'y}}, D., {Kipping}, D.~M., {Buchhave}, L.~A., et al. 2012, Science, 336, 1133

\bibitem[{{Omelyan} {et~al.}(2002){Omelyan}, {Mryglod}, \& {Folk}}]{2002O}
{Omelyan}, I.~P., {Mryglod}, I.~M., \& {Folk}, R. 2002, Computer Physics
  Communications, 146, 188
  
\bibitem[Petrovich et al.(2014)]{2014P} Petrovich, C., 
Tremaine, S., \& Rafikov, R.\ 2014, \apj, 786, 101 

\bibitem[Pinsonneault et al.(2012)]{2012P} Pinsonneault, 
M.~H., An, D., Molenda-{\.Z}akowicz, J., et al.\ 2012, \apjs, 199, 30 

\bibitem[Quillen(2008)]{2008Q} Quillen, A.~C.\ 2008, 
arXiv:0810.3679 

\bibitem[{{Rasio} \& {Ford}(1996)}]{1996R}
{Rasio}, F.~A., \& {Ford}, E.~B. 1996, Science, 274, 954

\bibitem[Rasmussen \& Williams(2006)]{rasmussen}
Rasmussen, C.~E., \& Williams, C.~K.~I.\ 2006,
Gaussian Processes for Machine Learning,
The MIT Press

\bibitem[{{Rauch} \& {Holman}(1999)}]{1999R}
{Rauch}, K.~P., \& {Holman}, M. 1999, \aj, 117, 1087

\bibitem[{{Rivera} {et~al.}(2010){Rivera}, {Laughlin}, {Butler}, {Vogt},
  {Haghighipour}, \& {Meschiari}}]{2010RL}
{Rivera}, E.~J., {Laughlin}, G., {Butler}, R.~P., et al. 2010, \apj, 719, 890

\bibitem[{{Sanchis-Ojeda} {et~al.}(2012){Sanchis-Ojeda}, {Fabrycky}, {Winn},
  {Barclay}, {Clarke}, {Ford}, {Fortney}, {Geary}, {Holman}, {Howard},
  {Jenkins}, {Koch}, {Lissauer}, {Marcy}, {Mullally}, {Ragozzine}, {Seader},
  {Still}, \& {Thompson}}]{2012SFW}
{Sanchis-Ojeda}, R., {Fabrycky}, D.~C., {Winn}, J.~N., et al. 2012, \nat, 487, 449

\bibitem[{{Sari} \& {Goldreich}(2004)}]{2004S}
{Sari}, R., \& {Goldreich}, P. 2004, \apjl, 606, L77

\bibitem[Smith et al.(2012)]{2012S} Smith, J.~C., Stumpe, 
M.~C., Van Cleve, J.~E., et al.\ 2012, \pasp, 124, 1000 

\bibitem[{{Socrates} {et~al.}(2012){Socrates}, {Katz}, {Dong}, \&
  {Tremaine}}]{2012SKDT}
{Socrates}, A., {Katz}, B., {Dong}, S., \& {Tremaine}, S. 2012, \apj, 750, 106

\bibitem[Stumpe et al.(2012)]{2012SS} Stumpe, M.~C., Smith, 
J.~C., Van Cleve, J.~E., et al.\ 2012, \pasp, 124, 985 

\bibitem[{{Takeda} {et~al.}(2007){Takeda}, {Ford}, {Sills}, {Rasio}, {Fischer},
  \& {Valenti}}]{2007T}
{Takeda}, G., {Ford}, E.~B., {Sills}, A., et al. 2007, \apjs, 168, 297

\bibitem[Takeda et al.(2008)]{2008T} Takeda, G., Kita, R., 
\& Rasio, F.~A.\ 2008, \apj, 683, 1063 

\bibitem[Teyssandier et al.(2013)]{2013T} Teyssandier, J., 
Naoz, S., Lizarraga, I., \& Rasio, F.~A.\ 2013, \apj, 779, 166 

\bibitem[Torres et al.(2012)]{2012T} Torres, G., Fischer, 
D.~A., Sozzetti, A., et al.\ 2012, \apj, 757, 161 

\bibitem[{{Tremaine} \& {Dong}(2012)}]{2012TD}
{Tremaine}, S., \& {Dong}, S. 2012, \aj, 143, 94

\bibitem[Tsang et al.(2014)]{2014T} Tsang, D., Turner, N.~J., 
\& Cumming, A.\ 2014, \apj, 782, 113 

\bibitem[Twicken et al.(2010)]{2010T} Twicken, J.~D., 
Chandrasekaran, H., Jenkins, J.~M., et al.\ 2010, \procspie, 7740,  

\bibitem[{{Valenti} \& {Fischer}(2005)}]{2005V}
{Valenti}, J.~A., \& {Fischer}, D.~A. 2005, \apjs, 159, 141

\bibitem[{{Valenti} \& {Piskunov}(1996)}]{1996V}
{Valenti}, J.~A., \& {Piskunov}, N. 1996, \aaps, 118, 595

\bibitem[{{Vogt} {et~al.}(1994){Vogt}, {Allen}, {Bigelow}, {Bresee}, {Brown},
  {Cantrall}, {Conrad}, {Couture}, {Delaney}, {Epps}, {Hilyard}, {Hilyard},
  {Horn}, {Jern}, {Kanto}, {Keane}, {Kibrick}, {Lewis}, {Osborne},
  {Pardeilhan}, {Pfister}, {Ricketts}, {Robinson}, {Stover}, {Tucker}, {Ward},
  \& {Wei}}]{1994V}
{Vogt}, S.~S., {Allen}, S.~L., {Bigelow}, B.~C., et al. 1994, in Society
  of Photo-Optical Instrumentation Engineers (SPIE) Conference Series, Vol.
  2198, Society of Photo-Optical Instrumentation Engineers (SPIE) Conference
  Series, ed. {D.~L.~Crawford \& E.~R.~Craine}, 362

\bibitem[{{Winn} {et~al.}(2010){Winn}, {Fabrycky}, {Albrecht}, \&
  {Johnson}}]{2010W}
{Winn}, J.~N., {Fabrycky}, D., {Albrecht}, S., \& {Johnson}, J.~A. 2010, \apjl,
  718, L145

\bibitem[{{Wu} \& {Lithwick}(2011)}]{2011WL}
{Wu}, Y., \& {Lithwick}, Y. 2011, \apj, 735, 109

\bibitem[{{Wu} \& {Lithwick}(2013)}]{2013WL}
---. 2013, \apj, 772, 74

\bibitem[{{Wu} \& {Murray}(2003)}]{2003W}
{Wu}, Y., \& {Murray}, N. 2003, \apj, 589, 605

\bibitem[Zhang et al.(2013)]{2013Z} Zhang, K., Hamilton, 
D.~P., \& Matsumura, S.\ 2013, \apj, 778, 6 

\end{thebibliography}

\end{document}